\documentclass[twocolumn,prb,superscriptaddress]{revtex4}
\usepackage[utf8]{luainputenc}
\usepackage{amsmath}
\usepackage{amssymb}
\usepackage{graphicx}
\usepackage{esint}

\makeatletter
\@ifundefined{textcolor}{}
{%
 \definecolor{BLACK}{gray}{0}
 \definecolor{WHITE}{gray}{1}
 \definecolor{RED}{rgb}{1,0,0}
 \definecolor{GREEN}{rgb}{0,1,0}
 \definecolor{BLUE}{rgb}{0,0,1}
 \definecolor{CYAN}{cmyk}{1,0,0,0}
 \definecolor{MAGENTA}{cmyk}{0,1,0,0}
 \definecolor{YELLOW}{cmyk}{0,0,1,0}
}

\@ifundefined{textcolor}{}{%
 \definecolor{BLACK}{gray}{0}
 \definecolor{WHITE}{gray}{1}
 \definecolor{RED}{rgb}{1,0,0}
 \definecolor{GREEN}{rgb}{0,1,0}
 \definecolor{BLUE}{rgb}{0,0,1}
 \definecolor{CYAN}{cmyk}{1,0,0,0}
 \definecolor{MAGENTA}{cmyk}{0,1,0,0}
 \definecolor{YELLOW}{cmyk}{0,0,1,0}
}

\newcommand{\rmc}{{\rm c}}
\newcommand{\rmd}{{\rm d}}
\newcommand{\rme}{{\rm e}}
\newcommand{\rmg}{{\rm g}}
\newcommand{\rmi}{{\rm i}}

\newcommand{\kk}{{\boldsymbol k}}
\newcommand{\rr}{{\boldsymbol r}}

\newcommand{\ca}{c^{\alpha}_{\boldsymbol k}}
\newcommand{\cb}{c^{\beta}_{\boldsymbol k}}

\newcommand{\hb}{{c^{\beta}_{-\boldsymbol k}}}

\makeatother

\begin{document}

\title{Photo-induced superconductivity in semiconductors}

\author{Garry Goldstein}

\affiliation{Department of Physics, Rutgers University, Piscataway, New Jersey
08854, USA}

\author{Camille Aron}

\affiliation{Department of Physics, Rutgers University, Piscataway, New Jersey
08854, USA}

\affiliation{Department of Electrical Engineering, Princeton University, Princeton,
New Jersey 08544, USA}

\author{Claudio Chamon}

\affiliation{Department of Physics, Boston University, Boston, Massachusetts 02215,
USA}
\begin{abstract}
We show that optically pumped \emph{semiconductors} can exhibit superconductivity.
We illustrate this phenomenon in the case of a two-band semiconductor
tunnel-coupled to broad-band reservoirs and driven by a continuous
wave laser. More realistically, we also show that superconductivity
can be induced in a two-band semiconductor interacting with a broad-spectrum
light source. We furthermore discuss the case of a three-band model
in which the middle band replaces the broad-band reservoirs as the
source of dissipation. In all three cases, we derive the simple conditions
on the band structure, electron-electron interaction, and hybridization
to the reservoirs that enable superconductivity. We compute the finite
superconducting pairing and argue that the mechanism can be induced
through both attractive and repulsive interactions and is robust to
high temperatures. 
\end{abstract}
\maketitle

\section{\label{sec:Introduction}Introduction}

Superconductivity is unarguably a fascinating phase of matter with
tremendous applications. This low-temperature instability towards
zero-resistivity corresponds to the emergence and the condensation
of Cooper pairs of electrons. In most simple metallic systems, the
pairing is achieved by phonon-mediated interactions~\cite{key-17}
and the superconducting temperature does not exceed a few Kelvin.
The last fifty years have seen some remarkable progress in the understanding
of superconductivity. Cuprates~\cite{key-33} and iron pnictides~\cite{key-34,key-35}
now offer critical temperatures on the order of a hundred Kelvin.
They were dubbed ``high-temperature superconductors'' as such temperatures
can be easily achieved with liquid nitrogen. All this allowed superconductivity
to become a cornerstone to many modern technological developments
\cite{key-17}. The Josephson effect is routinely used in superconducting
quantum interference devices (SQUIDs) \cite{key-17}, and its inherent
non-linearity is widely used to build qubits~\cite{key-36,key-37}.
The Meissner effect and the zero resistivity are used to realize powerful
magnets \cite{key-38}. However, the search for room-temperature superconductivity
is still a very active field of research \cite{key-39}.

Pioneering examples of the use of AC microwave fields in condensed
matter systems have been to enhance the critical temperature of regular
superconductors by redistributing the quasiparticle density near the
Fermi surface~\cite{key-17}. More recently, it was established that
an AC electric field leads to a renormalization of the lattice hopping
parameters~\cite{key-18,key-19,key-20}. It has been suggested that
in interacting systems such as the Bose-Hubbard model it is thereby
possible to induce a superfluid Mott insulator phase transition~\cite{key-21,key-22,key-23}.
Reversing the signs of the hoppings in a lattice model could be used
to realize frustrated classical spin systems~\cite{key-24}. In the
case of electrons driven by a laser field, many interesting phenomena
have been proposed~\cite{key-25,key-26}. These include dynamical
band flipping and splitting~\cite{key-27}, interaction strength
renormalization, changes in the sign of the effective interaction
strength leading to s-wave superconductivity with repulsive bare interactions
and negative absolute temperatures for a laser-driven band model.

In this work, we envision a novel route to achieve superconductivity
which consists of optically driving a two-band \emph{semiconductor}
to a suitable non-equilibrium steady state which supports \textit{interband}
pairing between electrons in the valence and conduction electrons.
Importantly, we shall demonstrate the robustness of this mechanism
with respect to temperature, up to room temperature, as long as it
is smaller than the semiconducting gap.

We note that the possibility of inducing superconductivity in a two
band model has been discussed in narrow, indirect gap semiconductors.~\cite{key-1,key-2,key-3,key-4,key-5}
However the mechanism proposed here is significantly different in
that it involves \textit{interband} pairing for wide gap semiconductors,
instead of \textit{intraband} pairing for narrow gap semiconductors.
Furthermore, in our mechanism the majority of the pairing occurs around
a resonant surface ${\cal S}_{\omega_{0}}$ (see Section~II) which
is not directly related to the band edge. Among the chief consequences
of the difference in pairing channel and its $k$-space location is
the fact that in our mechanism the pairing amplitude does not need
to be larger than the semiconducting gap in order to establish non-equilibrium
steady-state superconductivity, therefore making pairing more easily
attainable.

\begin{figure}
\begin{centering}
\includegraphics[angle=90,scale=0.3]{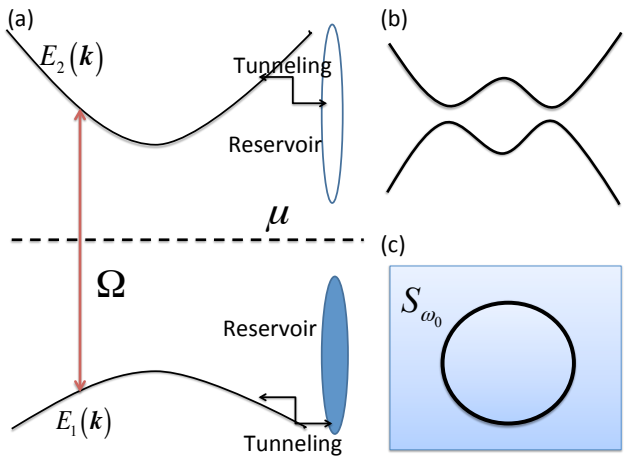} 
\par\end{centering}

\protect\protect\protect\protect\protect\caption{\label{fig:Energy_Levels-1}Energy levels and laser. (a) A continuous
wave laser drives transitions between the lower (1) and upper (2)
bands. Both bands are coupled to reservoirs. The chemical potentials
of the reservoirs $\mu$ is set in the gap. (b) In the rotating frame,
the laser induces an avoided band crossing (with splitting $\sim\left|\Omega\right|$).
(c) This creates an effective resonant surface ${\cal S}_{\omega_{0}}$
consisting of the set of momenta $\kk_{0}$ for which the laser resonantly
connects the two bands.}
\end{figure}

In Sect.~II, we take a pedagogical route to demonstrate this effect
by considering a model of a two-band semiconductor in tunneling contact
with two reservoirs provided, say, by a metallic plate {[}see Fig.~(\ref{fig:Energy_Levels-1}){]}.
We carefully show that it is possible to induce superconductivity
in this system under favorable conditions involving the electronic
dispersion, the electron-electron interaction, and the hybridizations
to the reservoirs as well as the chemical potential. To support the
validity of our analytical results in the steady-state, we perform
an exact numerical integration of the time dynamics. We show in particular
that the predicted non-trivial steady state is indeed reached dynamically.

In Sect.~III, we argue that the previous case can be reduced to a
simpler yet more realistic model of a two-band semiconductor -- not
in strong tunneling contact with any engineered external reservoirs
-- which is optically pumped by a broad-band light source. In many
ways, it is the most relevant model discussed in this manuscript and
the reader eager to learn about these results can directly jump to
Sect.~III which is written in a self-contained fashion.

In Sect.~IV, we provide an alternative derivation of the previous
results by means of a Keldysh formalism approach. In particular, this
allows to justify properly an approximation used to self-consistently
compute the superconducting pairing.

For the sake of completeness, we review in the Appendix the case of
a three-band semiconductor in which the extra band plays the role
of the reservoirs in Sect~II.

\section{\label{sec:Mean-field-Hamiltonian} Laser-driven dissipative two-band
semiconductor }

Let us consider a semiconductor with two relevant electronic bands:
the lower band ($\alpha=1$) with dispersion $E_{1}(\kk)$ and the
upper band ($\alpha=2$) with dispersion $E_{2}(\kk)$ are separated
by a gap $E_{\rmg}$. For the sake of simplicity, let us assume that
the dispersion is symmetric so that $E_{\alpha}\left(\kk\right)=E_{\alpha}\left(-\kk\right)$
for both bands $\alpha=1,2$; this will allow for s-wave superconductivity
without any energy mismatches. The semiconductor is driven by a continuous
coherent laser source with frequency $\omega_{0}$. This induces transitions
between the bands if there are momenta $\kk_{0}$ such that $E_{2}\left(\kk_{0}\right)=E_{1}\left(\kk_{0}\right)+\omega_{0}$.
In practice this condition is easily met and the corresponding momenta
lie on a finite closed surface ${\cal S}_{\omega_{0}}$ of the Brillouin
zone. In particular we assume that the level surfaces of $E_{1}\left(\kk_{0}\right)$
and $E_{2}\left(\kk_{0}\right)$ have good overlap (which would happen
for say parabolic bands). The laser acts as a source of energy and
we provide a heat sink by coupling each band to an independent reservoir
which can exchange particles and energy. Both reservoirs are kept
in equilibrium at temperature $T$ and chemical potential $\mu$.
In principle, one can also consider a single reservoir provided that
its density of state is broad enough to overlap with the upper and
lower bands. We set the chemical potential in the gap, exactly halfway
between the two bands, $\mu=[E_{2}\left(\kk_{0}\right)+E_{1}\left(\kk_{0}\right)]/2$.
This ensures that all quasiparticles have zero energy. In the rotating
frame, this will correspond to a zero-energy condition for the electrons
at $\kk_{0}\in{\cal S}_{\omega_{0}}$. Below, we measure energies
relative to $\mu$, \textit{i.e.} we set $\mu=0$. We shall see later
that the ability for the electronic bands to acquire non-trivial populations
is crucial to the occurrence of superconductivity. In the case at
hand, this is favored if the two reservoirs have different density
of states or different coupling strength to the bands. In the Appendix,
we shall see that a third band, or alternatively as in Sect.~\ref{sec:Optical-Pumping},
other $\kk$ modes in the same band can also play this role.

In order to establish that superconductivity can be realized in such
a driven-dissipative system, we first solve for its non-equilibrium
steady-state dynamics by means of a Master equation approach. Within
a self-consistent mean-field approach, we then obtain the criteria
for superconducting pairing and estimate the size of the superconducting
gap. Finally, we discuss the robustness of our results, in particular
against finite temperature.

\subsection{\label{sub:Mean-field-Hamiltonian-and}Mean-field Hamiltonian and
Master Equation}

We decompose the total model Hamiltonian into a laser-driven semiconductor
part (the system), a reservoir part (the bath), and a system-reservoir
coupling part: 
\begin{equation}
H=H_{{\rm sys}}+H_{{\rm bath}}+H_{{\rm sys-bath}}\label{eq:Main_Hamiltonian}
\end{equation}
where 
\begin{align}
H_{{\rm sys}}= & \sum_{\kk,\alpha}\; E_{\alpha}(\kk)\;{\ca}^{\dagger}\,\ca+\Omega(t)\,\sum_{\kk,\alpha,\beta}\;{\ca}^{\dagger}\,\sigma_{\alpha\beta}^{x}\,\cb\\
 & +\frac{\rmi}{2}\Delta\sum_{\kk,\alpha,\beta}\;{\ca}^{\dagger}\,\sigma_{\alpha\beta}^{y}\,{\hb}^{\!\!\!\!\dagger}-\frac{\rmi}{2}\Delta^{*}\sum_{\kk,\alpha,\beta}\;{\ca}\,\sigma_{\alpha\beta}^{y}\,{\hb},\nonumber \\
H_{{\rm bath}}= & \sum_{\kk,n,\alpha}\;\omega_{n}^{\alpha}(\kk)\;{a_{\kk,n}^{\alpha}}^{\!\!\!\!\dagger}\;{a_{\kk,n}^{\alpha}}\;,
\end{align}
and 
\begin{align}
H_{{\rm sys-bath}}=\sum_{\kk,n,\alpha}t_{\alpha}(\kk)\;\left[{\ca}^{\dagger}{a_{\kk,n}^{\alpha}}+{a_{\kk,n}^{\alpha}}^{\!\!\!\!\dagger}\;{\ca}\right]\;.
\end{align}
${\ca}$ and ${\ca}^{\dagger}$ are the creation and annihilation
operators of electrons with a quasi-momentum $\kk$ in the $\alpha$
band, $\alpha=1,2$. $\Omega\left(t\right)=\Omega\cos(\omega_{0}t)$
is the laser drive, and $\sigma^{x,y,z}$ are the usual Pauli matrices
acting on the band indices. The last two terms in $H_{{\rm sys}}$
originate from a microscopic electron-electron interaction which we
treat at a mean-field level (see also Sect.~II~b). $\Delta$ is
the complex order parameter which quantifies the superconducting pairing
between the bands and that will be determined self-consistently. The
$a_{\kk,n}^{\alpha}$'s represent the degrees of freedom of the non-interacting
reservoirs with energy $\omega_{n}^{\alpha}$. Here $n$ is a mode
label. We shall assume that the reservoirs have continuous density
of states given by $\nu_{\alpha}(\omega)$ and that they are weakly
coupled to the system, \textit{i.e.} $\left|t_{\alpha}^{2}\right|\nu_{\alpha}\ll E_{2}-E_{1}$
\cite{key-41}. In this case, the dynamics of the reduced density
matrix of the system, $\rho_{{\rm sys}}$, can be described by a Master
equation reading~\cite{key-1-1} \begin{widetext} 
\begin{align}
\frac{\rmd}{\rmd t}\,\rho_{{\rm sys}}= & -\rmi\left[H_{{\rm sys}},\rho_{{\rm sys}}\right]+\sum_{\kk,\alpha}\Gamma_{\alpha}\left(\kk\right)\left[n_{F}\left(E_{\alpha}\left(\kk\right)\right)\mathcal{D}[c_{\kk}^{\alpha\dagger}]\rho_{{\rm sys}}+\left(1-n_{F}\left(E_{\alpha}\left(\kk\right)\right)\right)\mathcal{D}[c_{\kk}^{\alpha}]\rho_{{\rm sys}}\right]\;,
\end{align}
where $n_{{\rm F}}(\epsilon)\equiv[1+\exp(\epsilon/T)]^{-1}$ is the
Fermi-Dirac distribution function, and the rates $\Gamma_{\alpha}(\kk)\equiv\pi\left|t_{\alpha}(\kk)\right|^{2}\nu_{\alpha}\big(E_{\alpha}(\kk)\big)$
are given by Fermi's golden rule. We note that for some decaying mechanisms
such as phonons not considered here, the rates $\Gamma_{1}$ and $\Gamma_{2}$
may be temperature dependent. The Lindblad-type dissipators are defined
as $\mathcal{D}[X]\rho\equiv\left(X\rho X^{\dagger}-X^{\dagger}X\rho+{\rm {h.c.}}\right)/2$.
We neglected the Lamb-shift corrections (real part of hybridization
self-energy). We may now write the equations of motion for the populations,
coherences and anomalous correlators $n_{\kk}^{\alpha\beta}\equiv\langle{\ca}^{\dagger}\,{\cb}\rangle$
and $s_{\kk}^{\alpha\beta}\equiv\langle{\ca}^{\dagger}\,{\hb}^{\!\!\!\!\dagger}\,\rangle$
with $\alpha,\beta=1,2$: \begin{subequations} \label{eq:Equationsofmotion}
\begin{eqnarray}
\frac{\rmd}{\rmd t}n_{\kk}^{11} & = & -\rmi\Omega(t)\left(n_{\kk}^{12}-n_{\kk}^{21}\right)+\rmi\Delta\, s_{\kk}^{21}-\rmi\Delta^{*}\,{s_{\kk}^{21}}^{*}-2\Gamma_{1}(\kk)\left[n_{\kk}^{11}-n_{{\rm F}}\big(E_{1}(\kk)\big)\right],\\
\frac{\rmd}{\rmd t}n_{\kk}^{22} & = & -\rmi\Omega(t)\left(n_{\kk}^{21}-n_{\kk}^{12}\right)+\rmi\Delta\, s_{\kk}^{21}-\rmi\Delta^{*}\,{s_{\kk}^{21}}^{*}-2\Gamma_{2}(\kk)\left[n_{\kk}^{22}-n_{{\rm F}}\big(E_{2}(\kk)\big)\right],\\
\frac{\rmd}{\rmd t}n_{\kk}^{21} & = & \rmi[E_{2}(\kk)-E_{1}(\kk)\big]\, n_{\kk}^{21}-\rmi\Omega(t)\left(n_{\kk}^{22}-n_{\kk}^{11}\right)-[\Gamma_{2}(\kk)+\Gamma_{1}(\kk)]\, n_{\kk}^{21},\\
\frac{\rmd}{\rmd t}s_{\kk}^{21} & = & \rmi[E_{2}(\kk)+E_{1}(\kk)]\, s_{\kk}^{21}+\rmi\Delta^{*}\left(n_{\kk}^{11}+n_{\kk}^{22}-1\right)-[\Gamma_{2}(\kk)+\Gamma_{1}(\kk)]\, s_{\kk}^{21}\;,
\end{eqnarray}
\end{subequations}in which we made use of the identity ${\rm tr}\big(O\,\mathcal{D}[X]\rho\big)={\rm tr}\big([X^{\dagger},O]X\rho\big)+{\rm tr}\big(X^{\dagger}[O,X]\rho\big)=\langle[X^{\dagger},O]X\rangle_{\rho}+\langle X^{\dagger}[O,X]\rangle_{\rho}$
repeatedly. We then perform a rotating wave approximation (RWA) to
eliminate the explicit time dependence of these equations. This consists
in rotating all the operators of the theory with the unitary 
\begin{align}
U\equiv U_{c}\otimes U_{a}\;,
\end{align}
where 
\begin{align}
U_{c}\equiv\exp\left[\frac{\rmi}{2}\omega_{0}t\sum_{\kk}\left(c_{\kk}^{1\dagger}c_{\kk}^{1}-c_{\kk}^{2\dagger}c_{\kk}^{2}\right)\right]\mbox{ and }U_{a}\equiv\exp\left[\frac{\rmi}{2}\omega_{0}t\sum_{\kk,n}\left(a_{\kk,n}^{1\dagger}a_{\kk,n}^{1}-a_{\kk,n}^{2\dagger}a_{\kk,n}^{2}\right)\right]\;.
\end{align}
In particular, $c_{\kk}^{1}\mapsto\widetilde{c}_{\kk}^{1}=c_{\kk}^{1}\,\rme^{-\rmi\omega_{0}t/2}$,
$c_{\kk}^{2}\mapsto\widetilde{c}_{\kk}^{2}=c_{\kk}^{2}\,\rme^{\rmi\omega_{0}t/2}$,
and $H\mapsto\widetilde{H}=U\left[H-\rmi\partial_{t}\right]U^{\dagger}$
so that the energies are shifted to $\widetilde{E}_{1}(\kk)=E_{1}(\kk)+\omega_{0}/2$
and $\widetilde{E}_{2}(\kk)=E_{2}(\kk)-\omega_{0}/2$. Note that in
the rotating frame, $\widetilde{n}_{\kk}^{11}=n_{\kk}^{11}$, $\widetilde{n}_{\kk}^{22}=n_{\kk}^{22}$,
and $\widetilde{s}_{\kk}^{12}=s_{\kk}^{12}$ are invariant, but $\widetilde{n}_{\kk}^{12}=n_{\kk}^{12}\,\rme^{-\rmi\omega_{0}t}$
and $\widetilde{n}_{\kk}^{21}=n_{\kk}^{21}\,\rme^{\rmi\omega_{0}t}$.
We drop all terms rotating at $2\omega_{0}$ since they are not resonant
with any transition. We also drop the $\kk$-dependence of the decay
rates $\Gamma_{1,2}(\kk)\to\Gamma_{1,2}$, which is justified by assuming
their weak momentum dependence in the small window of momenta around
the surface of resonant condition ${\cal S}_{\omega_{0}}$. Altogether,
we obtain \begin{subequations} \label{eq:Equations_RWA} 
\begin{eqnarray}
\frac{\rmd}{\rmd t}\widetilde{n}_{\kk}^{11} & = & -\frac{\rmi}{2}\Omega\left(\widetilde{n}_{\kk}^{12}-\widetilde{n}_{\kk}^{21}\right)+\rmi\Delta\,{\widetilde{s}}_{\kk}^{21}-\rmi\Delta^{*}\,{\widetilde{s}}{_{\kk}^{21}}^{*}-2\Gamma_{1}\left[\widetilde{n}_{\kk}^{11}-n_{{\rm F}}\big(E_{1}(\kk)\big)\right]\;,\label{eq:n11_rotated}\\
\frac{\rmd}{\rmd t}\widetilde{n}_{\kk}^{22} & = & -\frac{\rmi}{2}\Omega\left(\widetilde{n}_{\kk}^{21}-\widetilde{n}_{\kk}^{12}\right)+\rmi\Delta\,{\widetilde{s}}_{\kk}^{21}-\rmi\Delta^{*}\,{\widetilde{s}}{_{\kk}^{21}}^{*}-2\Gamma_{2}\left[\widetilde{n}_{\kk}^{22}-n_{{\rm F}}\big(E_{2}(\kk)\big)\right]\;,\label{eq:n22_rotated}\\
\frac{\rmd}{\rmd t}\widetilde{n}_{\kk}^{21} & = & (\rmi\varepsilon_{\kk}-\Gamma)\,\widetilde{n}_{\kk}^{21}-\frac{\rmi}{2}\Omega\left(\widetilde{n}_{\kk}^{22}-\widetilde{n}_{\kk}^{11}\right)\;,\label{eq:n21_rotated}\\
\frac{\rmd}{\rmd t}{\widetilde{s}}_{\kk}^{21} & = & \left(\rmi E_{\kk}-\Gamma\right)\,{\widetilde{s}}_{\kk}^{21}+\rmi\Delta^{*}\left({\widetilde{n}}_{\kk}^{11}+{\widetilde{n}}_{\kk}^{22}-1\right)\;,\label{eq:s21_rotated}
\end{eqnarray}
\end{subequations} \end{widetext} where we defined $\Gamma\equiv\Gamma_{1}+\Gamma_{2}$,
$\varepsilon_{\kk}\equiv\widetilde{E}_{2}(\kk)-\widetilde{E}_{1}(\kk)$,
and $E_{\kk}\equiv\widetilde{E}_{2}(\kk)+\widetilde{E}_{1}(\kk)$.
Where we have used the symmetry between $\kk$ and $-\kk$ stemming
from $E_{\alpha}\left(\kk\right)=E_{\alpha}\left(-\kk\right)$. This
reduces all computations to just one wavevector $\kk$.

The steady-state values of populations, coherences and anomalous correlators
can now be solved by setting the left-hand side of Eqs.~(\ref{eq:Equations_RWA})
to zero. We find that 
\begin{equation}
{\widetilde{s}}_{\kk}^{21}=-\frac{\Delta^{*}}{E_{\kk}+\rmi\Gamma}\;\left({\widetilde{n}}_{\kk}^{11}+{\widetilde{n}}_{\kk}^{22}-1\right)\;,\label{eq:Delta-steady-state}
\end{equation}
where ${\widetilde{n}}_{\kk}^{11}+{\widetilde{n}}_{\kk}^{22}-1$ measures
the fraction of the total population that can be borrowed from, or
shifted to, the ``storage'' constituted by the reservoirs or by
the other $\kk$ modes away from resonance. It is given by 
\begin{align}
\widetilde{n}_{\kk}^{11}+\widetilde{n}_{\kk}^{22}\!-\!1\!\approx\!\frac{\gamma_{1}-\gamma_{2}}{\Xi}\frac{\Omega^{2}}{\epsilon_{\kk}^{2}+\Gamma^{2}}\left[n_{F}\left(E_{1}\!\left(\kk\right)\right)\!-\! n_{F}\!\left(E_{2}\left(\kk\right)\right)\right],\label{eq:population}
\end{align}
where we defined $\gamma_{1,2}\equiv\Gamma_{1,2}/\Gamma$, 
\begin{equation}
\Xi\equiv4\gamma_{1}\gamma_{2}+\frac{4|\Delta|^{2}}{{E_{\kk}}^{2}+\Gamma^{2}}+\frac{\Omega^{2}}{\epsilon_{\kk}^{2}+\Gamma^{2}}\left[1+\frac{4|\Delta|^{2}}{{E_{\kk}}^{2}+\Gamma^{2}}\right]\;,
\end{equation}
and we neglected a term proportional to $n_{{\rm F}}\left(E_{1}(\kk)\right)+n_{{\rm F}}\left(E_{2}(\kk)\right)-1$
since this factor vanishes at zero temperature and is exponentially
suppressed for temperatures smaller than the semiconducting gap $E_{\rmg}$.
We note that both a large $\gamma_{1}\gamma_{2}$ and a large $\left|\Delta\right|$
lead to a decrease in $\widetilde{n}_{\kk}^{11}+\widetilde{n}_{\kk}^{22}\!-\!1$.

Anticipating what follows, we shall see that only a non-vanishing
value of ${\widetilde{n}}_{\kk}^{11}+{\widetilde{n}}_{\kk}^{22}-1$,
\textit{i.e.} a finite population deviation from the equilibrium situation,
will amount to superconductivity. It is quite transparent from Eq.~(\ref{eq:population})
that in order to obtain such non-trivial band populations, one must
drive the system ($\Omega\neq0$) and the decay rates $\Gamma_{1}$
and $\Gamma_{2}$ must be different ($\gamma_{1}\neq\gamma_{2}$).

When the drive $\Omega$ is large compared to $\Gamma$, the ratio
$\Omega^{2}/(\epsilon_{\kk}^{2}+\Gamma^{2})$ is very large near the
resonance ($\epsilon_{\kk}=0$). In this case, and when the temperature
is much smaller than the semiconducting gap, the non-equilibrium population
deviation simplifies to 
\begin{equation}
\widetilde{n}_{\kk}^{11}+\widetilde{n}_{\kk}^{22}-1\approx\frac{E_{\kk}^{2}+\Gamma^{2}}{E_{\kk}^{2}+\Gamma^{2}+4|\Delta|^{2}}\;(\gamma_{1}-\gamma_{2})\;,\label{eq:deviation_large_omega}
\end{equation}
which holds in a range of width $\Omega$ near the resonance. We note
that this approximation is also valid for moderate $\Omega$ when
$\gamma_{1}\gamma_{2}\cong0$ and $\Delta$ is small. Hence, $\Omega$
plays the role of a cut-off, and for energies $|\epsilon_{\kk}|<\Omega$
we can use the approximate expression in the equation above. Notice
that one can achieve finite non-equilibrium population deviations
in this range of $\epsilon_{\kk}$, on the order of $\gamma_{1}-\gamma_{2}$.
Moreover, notice that the sign of this deviation depends on which
of the decay rates $\Gamma_{1}$ or $\Gamma_{2}$ is larger, see also
Fig.~(2).

In the opposite case in which the decay rate $\Gamma$ is much larger
than the drive $\Omega$, the non-equilibrium population deviation
is (for bath temperatures much smaller than the semiconducting gap)
\begin{equation}
\widetilde{n}_{\kk}^{11}+\widetilde{n}_{\kk}^{22}-1\approx\frac{{E_{\kk}}^{2}+\Gamma^{2}}{{E_{\kk}}^{2}+\Gamma^{2}+\frac{\;|\Delta|^{2}}{\gamma_{1}\gamma_{2}}}\;\frac{\gamma_{1}-\gamma_{2}}{4\gamma_{1}\gamma_{2}}\;\frac{\Omega^{2}}{\epsilon_{\kk}^{2}+\Gamma^{2}}\;.\label{eq:Population_deviation}
\end{equation}

\begin{figure}
\begin{centering}
\includegraphics[angle=90,scale=0.25]{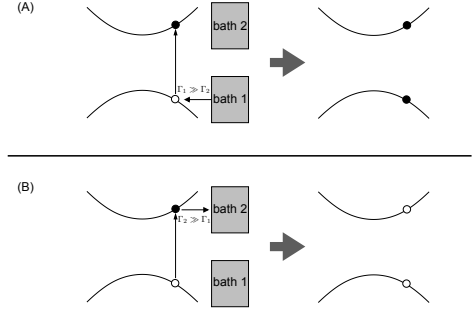} 
\par\end{centering}

\protect\protect\protect\protect\protect\caption{\label{fig:population_figure}Non-equilibrium population deviation
due to driving and dissipation. The laser causes an electron in the
valance band to transition into the conduction band. The figure illustrates
particular examples when the two rates $\Gamma_{1,2}$ differ substantially.
In (A), the rate $\Gamma_{1}\gg\Gamma_{2}$, so the reservoirs fill
the hole in the valance band much faster than the electron in the
conduction band can relax back; the state is blocked, and one has
$n_{\kk}^{11}+n_{\kk}^{22}-1\sim+1$. In (B), the rate $\Gamma_{2}\gg\Gamma_{1}$,
so the reservoirs remove the electron in the conduction band much
faster than it can relax back; one is left with two holes, and one
has $n_{\kk}^{11}+n_{\kk}^{22}-1\sim-1$ in this example.}
\end{figure}

\subsection{\label{sec:Self-Consistency-Equation}Self-Consistency Equation}

We now solve self-consistently for the superconducting gap. The pairing
part of the mean-field Hamiltonian originates from a microscopic Hamiltonian
which involves a density-density type of interaction between the electrons
in the semiconductor. The mean-field decoupling for this microscopic
interaction of strength $V$ (in a system of volume ${\cal V}$) is
given by: 
\begin{eqnarray}
H_{e-e} & = & \frac{1}{{\cal V}}\sum_{\kk,\kk'}\; V\;{c_{\mathbf{k}}^{2}}^{\dagger}\,{c_{-\mathbf{k}}^{1}}^{\!\!\!\dagger}\;\;{c_{\mathbf{k}'}^{1}}\,{c_{-\mathbf{k}'}^{2}}\label{eq:superconductivity-1}\\
\qquad & \rightarrow & \sum_{\kk}\left(\Delta\;{c_{\mathbf{k}}^{2}}^{\dagger}\,{c_{-\mathbf{k}}^{1}}^{\!\!\!\dagger}+\Delta^{*}\;{c_{\mathbf{k}}^{1}}\,{c_{-\mathbf{k}}^{2}}\right)\;,
\end{eqnarray}
with 
\begin{equation}
\Delta^{*}=\frac{1}{{\cal V}}\sum_{\kk}V\langle{c_{\mathbf{k}}^{2}}^{\dagger}\,{c_{-\mathbf{k}}^{1}}^{\!\!\!\dagger}\rangle\xrightarrow[{\cal V}\to\infty]{}\int(\rmd{\kk})\; V\langle{c_{\mathbf{k}}^{2}}^{\dagger}{c_{-\mathbf{k}}^{1}}^{\!\!\!\dagger}\rangle\;,\label{eq:self-consistent-condition}
\end{equation}
where we wrote $(\rmd\kk)\equiv\rmd^{d}\kk/(2\pi)^{d}$ to shorten
notations.

Eq.~(\ref{eq:self-consistent-condition}) is solved self-consistently
by using the anomalous correlator in Eq.~(\ref{eq:Delta-steady-state}).
The correct self-consistent condition involves only the real part
of Eq.~(\ref{eq:Delta-steady-state}); this assertion will be justified
in Sect.~\ref{sec:Keldysh-Calculation} where we properly obtain
the self-consistency relation from a saddle point condition (notice
that this is trivially true in the limit $\Gamma\to0$). More precisely
we use the self consistency relation: 
\begin{equation}
\Delta^{*}=\int(\rmd{\kk})\; V\tilde{s}_{\kk}^{21}Re\left(\frac{1}{E_{\kk}+i\Gamma}\right)\cdot\left(E_{\kk}+i\Gamma\right)\label{eq:self_consistency_corrected}
\end{equation}
We derive this rigorously in Sect.~\ref{sec:Keldysh-Calculation}.
Assuming that $\Omega\gg\Gamma$ and using Eq.~(\ref{eq:deviation_large_omega})
for the populations, the resulting gap equation reads 
\begin{eqnarray}
1 & = & -V\;\int(\rmd\kk)\;\frac{E_{\kk}}{E_{\kk}^{2}+\Gamma^{2}+4\left|\Delta\right|^{2}}\;(\gamma_{1}-\gamma_{2})\label{eq:gap-equation}\\
 & = & -N_{0}V\,(\gamma_{1}-\gamma_{2})\;\int_{-\Omega}^{\Omega}\!\!\!\ \rmd\epsilon\;\frac{E(\epsilon)}{E^{2}(\epsilon)+4|\Delta|^{2}+\Gamma^{2}}\;,\nonumber 
\end{eqnarray}
with $N_{0}\equiv\int(\rmd\kk)\,\delta(\epsilon_{\kk})$ the density
of states near the resonance.

Below, we study the solutions of the self-consistent equation in a
few relevant cases.

\subsubsection{Bands with opposite velocities at resonance}

This is a very favorable case, so let us start with it. On the resonant
surface ${\cal S}_{\omega_{0}}$, the dispersion relations of both
bands can be Taylor-expanded as $\widetilde{E}_{1,2}=v_{1,2}\, q_{\perp}+\kappa_{1,2}\, q_{\perp}^{2}+\dots$,
where $q_{\perp}$ is the momentum perpendicular to ${\cal S}_{\omega_{0}}$.
So $\epsilon=v_{-}\, q_{\perp}+\kappa_{-}\, q_{\perp}^{2}+\dots$
and $E=v_{+}\, q_{\perp}+\kappa_{+}\, q_{\perp}^{2}+\dots$, where
$v_{\pm}\equiv v_{2}\pm v_{1}$ and $\kappa_{\pm}\equiv\kappa_{2}\pm\kappa_{1}$.
If the velocities are opposite in the two bands, \textit{i.e.} $v_{+}=0$,
one can express $E(\epsilon)\approx(\kappa_{+}/v_{-}^{2})\;\epsilon^{2}$.
Upon using this $E(\epsilon)$ in Eq.~(\ref{eq:gap-equation}) and
extending the limits of integration in Eq.~(\ref{eq:gap-equation})
to $\pm\infty$ (for large $\Omega$), we obtain 
\begin{eqnarray}
1=-\frac{\pi}{\sqrt{2}}N_{0}V\,\frac{|v_{-}|\,{\rm sgn}\,\kappa_{+}}{\sqrt{|\kappa_{+}|}}\,\frac{\gamma_{1}-\gamma_{2}}{(4|\Delta|^{2}+\Gamma^{2})^{1/4}}\;.
\end{eqnarray}
We note that to get exact results in Eqs.~(\ref{eq:Delta-steady-state})
and (\ref{eq:population}). Notice that this equation can be satisfied
for \emph{both attractive or repulsive interactions} depending on
the relative signs of $\gamma_{1}-\gamma_{2}$ and of $\kappa_{+}$.
Superconductivity is possible if the sign of $V$ satisfies 
\begin{eqnarray}
{\rm sgn}\, V={\rm sgn}(\gamma_{2}-\gamma_{1})\times{\rm sgn}\,\kappa_{+}\;,\label{eq:signcond}
\end{eqnarray}
and its magnitude satisfies the threshold condition 
\begin{eqnarray}
|V|\ge V_{\rmc}\equiv\frac{\sqrt{2}}{\pi}\frac{1}{N_{0}}\frac{\sqrt{|\kappa_{+}|}}{|v_{-}|}\frac{\sqrt{\Gamma}}{|\gamma_{1}-\gamma_{2}|}\,\;.\label{eq:threshold}
\end{eqnarray}
This expresses the fact that superconductivity is favored by small
and different decay rates.

If the conditions in Eqs.~(\ref{eq:signcond}) and (\ref{eq:threshold})
are met, the superconducting gap is given by 
\begin{eqnarray}
|\Delta|=\frac{\Gamma}{2}\;\sqrt{\left(\frac{V}{V_{\rmc}}\right)^{4}-1}\;.\label{eq:delta_large_omega}
\end{eqnarray}
For large coupling constant, the gap scales as the square of the interaction
strength $V$. Notice also that the gap does not vanish in the limit
$\Gamma\to0$ because the threshold disappears simultaneously. In
this limit, 
\begin{eqnarray}
|\Delta|\xrightarrow[\Gamma\to0]{}\frac{\pi^{2}}{4}\left(N_{0}V\right)^{2}\,\frac{v_{-}^{2}}{|\kappa_{+}|}\,(\gamma_{1}-\gamma_{2})^{2}\;.
\end{eqnarray}

\paragraph{Robustness.}

\begin{figure}
\begin{centering}
\includegraphics[angle=90,scale=0.35]{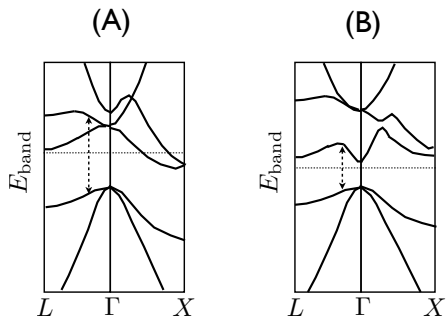} 
\par\end{centering}

\protect \protect\protect\protect\protect\protect\protect\protect\protect\caption{\label{fig:fig:extremum} Examples of how to choose optimal conditions.
One must seek points in the Brillouin zone where two bands have opposite
velocities. The transitions are depicted by the vertical dashed line,
whose length determines the laser frequency $\omega_{0}$. Notice
that the transition of choice does not need to be between two consecutive
bands, as is the case depicted in (A). In (B), the transition of choice
is between two consecutive bands. The horizontal dashed line demarcates
the position of the chemical potential, which can be chosen by doping
or gating. The topologies of the band structures were sketched to
resemble the bands in Si (A) and in GaAs (B).}
\end{figure}

Let us examine the domain of validity of our results. Let us first
argue that the condition $v_{+}=0$ that we used above can be achieved
by a proper choice of the laser frequency $\omega_{0}$. In practice,
one may proceed as follows. The resonance surface ${\cal S}_{\omega_{0}}$
can be swept as one changes $\omega_{0}$. At $\kk_{0}\in{\cal S}_{\omega_{0}}$,
$\epsilon_{\kk_{0}}=0$ by definition. Assume for simplicity a spherical-symmetric
dispersion. As one scans $\omega_{0}$, one should search for the
frequency for which $E_{\kk_{0}}$ reaches an extremum, either a minimum
or maximum. The extremum would correspond to a zero of $v_{+}$. Finding
the extremum condition may require using higher and lower bands; we
illustrate this for a few examples of band structure topologies in
Fig.~(\ref{fig:fig:extremum}). By changing the chemical potential,
one can make the value of the extremum be zero, \textit{i.e.} $E_{\kk_{0}}=0$,
and therefore $E(\epsilon)\propto\epsilon^{2}$.

Additionally, we note that our results are relatively stable in the
case of a non-vanishing $v_{+}$. Indeed, our results are essentially
unchanged as long as 
\begin{equation}
\left|v_{+}\right|\leq\sqrt{|\kappa_{+}|\;\Gamma}\;\frac{\left|V\right|}{V_{\rmc}}\;.\label{eq:v+inequality-1}
\end{equation}

Most importantly, our results are robust against finite temperatures
of the reservoirs. Indeed, this corresponds to changes in $n_{{\rm F}}(E_{1}(\kk))$
and $n_{{\rm F}}(E_{2}(\kk))$ which may be neglected for temperatures
less then the semiconducting gap $E_{\rmg}$.

\subsubsection{\label{sub:Weak-Rabi-frequency}Weak Rabi frequency}

Previously we considered the case in which the laser Rabi frequency
$\Omega$ was large compared to the decay rate $\Gamma$. This condition
is most favorable towards superconducting pairing; however for many
systems it is not satisfied. For lasers with moderate power (say on
the order of milliwatts) and semiconductors at room temperatures,
the laser Rabi frequency is several hundred megahertz while the carrier
decay rate is several tens of gigahertz. It is therefore relevant
to repeat the previous analysis in the less favorable case in which
the Rabi frequency is less than the particle decay rate.

Following the steps of Sect.~II~B~1, but using here the non-equilibrium
population deviation given in Eq.~(\ref{eq:Population_deviation}),
the superconducting self-consistency equation now reads 
\begin{align}
1= & -\frac{\pi}{4\sqrt{2}}N_{0}V\frac{|v_{-}|\,{\rm sgn}\,\kappa_{+}\,\sqrt{\left|\kappa_{+}\right|}}{v_{-}^{2}\sqrt{|\Delta|^{2}+\gamma_{1}\gamma_{2}\Gamma^{2}}+\sqrt{\gamma_{1}\gamma_{2}}\Gamma^{2}|\kappa_{+}|}\nonumber \\
 & \quad\times\frac{1}{(\gamma_{1}\gamma_{2})^{1/4}}\,\frac{\gamma_{1}-\gamma_{2}}{\left(|\Delta|^{2}+\gamma_{1}\gamma_{2}\Gamma^{2}\right)^{1/4}}\,\Omega^{2}\;.\label{eq:self_consistency_small_omega}
\end{align}
We recover the previous condition on the sign of the electron-electron
interaction, namely 
\begin{eqnarray}
{\rm sgn}\, V={\rm sgn}(\gamma_{2}-\gamma_{1})\times{\rm sgn}\,\kappa_{+}\;,\label{eq:sign2}
\end{eqnarray}
and the threshold condition now reads 
\begin{eqnarray}
|V|\ge{V}'_{c} & \equiv & \frac{4\sqrt{2}}{\pi}\frac{1}{N_{0}}\;\frac{v_{-}^{2}+|\kappa_{+}|\Gamma}{|v_{-}|\,\sqrt{|\kappa_{+}|}}\frac{\gamma_{1}\gamma_{2}\,\Gamma^{3/2}}{|\gamma_{1}-\gamma_{2}|}\frac{1}{\Omega^{2}},\label{eq:thre2}
\end{eqnarray}
Compared to the case $\Omega\gg\Gamma$ {[}see Eq.~(\ref{eq:threshold}){]},
the threshold condition has changed by a factor $4\gamma_{1}\gamma_{2}(\Gamma^{2}/\Omega^{2})\times[\left|\kappa_{+}\right|\Gamma/(\left|v_{-}\right|^{2}+\Gamma\left|\kappa_{+}\right|)]$.
We note that, while in this case both factors $\Gamma^{2}/\Omega^{2}$
and $\left|\kappa_{+}\right|\Gamma/(\left|v_{-}\right|^{2}+\Gamma\left|\kappa_{+}\right|)$
increase the threshold, this could be compensated if the two bands
have rather different decay rates, in which case the factor $\gamma_{1}\gamma_{2}$
can be small. If both conditions in Eqs.~(\ref{eq:sign2}) and (\ref{eq:thre2})
are met and $\left|v_{-}\right|^{2}\gg\left|\kappa_{+}\right|\Gamma$,
the gap is then given by 
\begin{eqnarray}
|\Delta|=\Gamma\,\sqrt{\gamma_{1}\gamma_{2}}\,\sqrt{\left(\frac{|V|}{{V}_{\rmc}'}\right)^{4/3}-1}\;.\label{eq:gap-small-Rabi}
\end{eqnarray}
Note that for large electronic interactions $|V|$, the gap is linear
in the decay rate $\Gamma$.

\paragraph{Vanishing decay rates.}

One particularly interesting case is when the Rabi frequency is small
but the two decay rates $\Gamma_{1}$ and $\Gamma_{2}$ are very different
so that $\gamma_{1}\gamma_{2}\approx0$. Repeating the previous steps,
the superconducting self-consistency equation reads 
\begin{align}
1= & \frac{\pi}{\sqrt{2}}N_{0}V\,(\gamma_{2}-\gamma_{1})\,\Omega\nonumber \\
 & \times\frac{|v_{-}|\,{\rm sgn}\,\kappa_{+}}{\left[\Omega^{2}\left|\kappa_{+}\right|^{2}\Gamma^{2}(4|\Delta|^{2}+\Omega^{2})\right]^{1/4}+2\left|v_{-}\right|\left|\Delta\right|}\label{eq:self_consistency_small_omega-1}
\end{align}
yielding the condition 
\begin{eqnarray}
{\rm sgn}\, V={\rm sgn}(\gamma_{2}-\gamma_{1})\times{\rm sgn}\,\kappa_{+}\;,
\end{eqnarray}
and the threshold 
\begin{equation}
|V|\ge{V}_{\rmc}''\equiv\frac{\sqrt{2}}{\pi}\frac{1}{N_{0}}\;\frac{\sqrt{|\kappa_{+}|}}{|v_{-}|}\,\frac{\Gamma}{|\gamma_{1}-\gamma_{2}|}\;.\label{eq:supercondcuting_treshhold}
\end{equation}
We note that this is the same threshold as in Eq.~(\ref{eq:threshold})
where we considered the case of a large Rabi frequency $\Omega\gg\Gamma$.
Whenever this threshold is satisfied in the case of a large $\left|v_{-}\right|$,
the superconducting order parameter reads 
\begin{equation}
\left|\Delta\right|=\frac{\Omega}{2}\frac{\sqrt{\Gamma|\kappa_{+}|}}{\left|v_{-}\right|}\left(\frac{\left|V\right|}{{V}_{c}''}-1\right)\;.\label{eq:magnitude_order_parameter}
\end{equation}

\subsection{\label{sec:Numeric_Simulations}Dynamics of the order parameter}

\begin{figure}
\begin{centering}
\vspace{5cm}
\hspace{-2cm}\includegraphics[bb=100bp 0bp 360bp 252bp,scale=0.4]{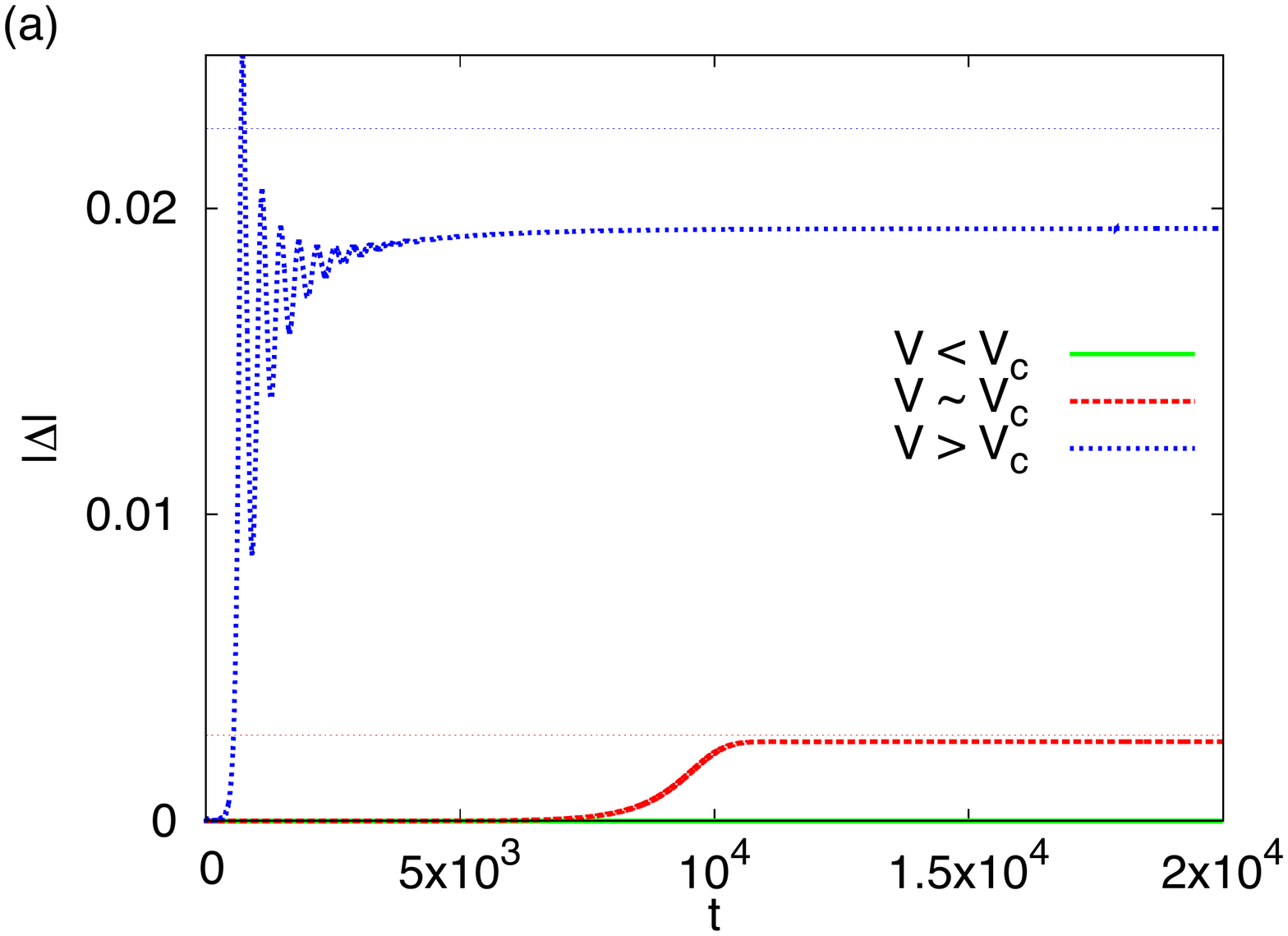}
\vspace{2.5cm}

\par\end{centering}

\begin{centering}
\hspace{-2cm}\includegraphics[bb=100bp 0bp 360bp 252bp,scale=0.4]{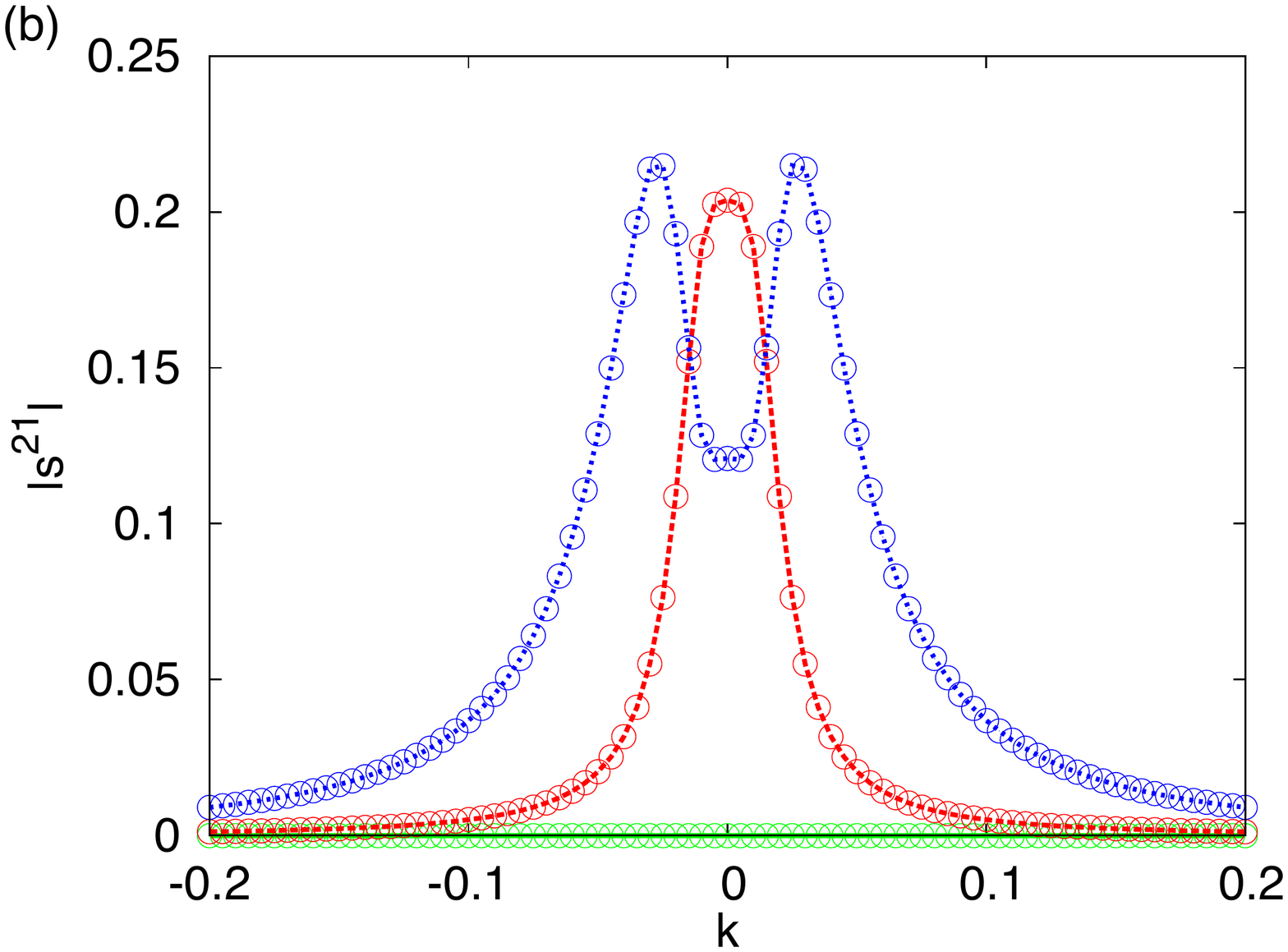}
\vspace{-3cm}

\par\end{centering}

\protect \protect\protect\protect\protect\protect\protect\protect\protect\caption{\label{fig:Good_match_plots}(a) Time evolution of the order parameter
$\Delta$ for three representative values of the coupling constant:
$V=5>V_{c}=2.14$, $V=2.3\sim V_{c}$ and $V=1<V_{c}$. The straight
lines correspond to the analytic expressions in Eq.~(\ref{eq:delta_large_omega}).
Here $V_{c}$ is computed using Eq.~(\ref{eq:threshold}) (corrected
to account for cutoff effects). (b) Perfect matching between the steady-state
anomalous correlator ${\widetilde{s}}_{\kk}^{21}$ as given by Eq.~(\ref{eq:Delta-steady-state})
(straight lines) and the anomalous correlator ${\widetilde{s}}_{\kk}^{21}$
obtained numerically after the time dynamics have converged (circles).
The color coding is the same as in (a). ($\kappa_{+}=-50$, $v_{-}=10$,
$\Omega=0.5$, $\Gamma=10^{-2}$ and $\gamma_{2}=10^{-3}$). We note
that in a typical semiconductor $\Gamma\approx10^{-2}eV$. Using this
value for $\Gamma$ we obtain that time is measure in units of $6.6\times10^{-16}$s
and $\Delta$ is measured in units of 1eV. Also for typical semiconductors
$v_{-}=10^{-2}c$, so the momentum is measured in unites of $10^{3}eV/c\approx5.3\cdot10^{25}kg\cdot m/s$.
Furthermore using these values the coupling constant $V$ is measured
in unites of $0.14N_{0}^{-1}$ (where $N_{0}$ is the density of states
at the surface ${\cal S}_{\omega_{0}}$).}
\end{figure}

We now confirm our analytic predictions by numeric integration of
the equations of motion, see Eqs.~(\ref{eq:n11_rotated}), (\ref{eq:n22_rotated}),
(\ref{eq:n21_rotated}) and (\ref{eq:s21_rotated}). We start by briefly
describing our numerical simulation procedure. For simplicity we consider
the case when the $E_{\alpha}\left(\kk\right)$ are spherically symmetric.
Furthermore by focusing on the region near the resonant surface ${\cal S}_{\omega_{0}}$
we may ignore variations in the density of states. In this case, within
mean field, we may reduce the dynamics of the 3-d model to the dynamics
of an equivalent one dimensional model where for simplicity we can
mathematically shift the surface ${\cal S}_{\omega_{0}}$ to the wavevector
$\kk_{0}=0$. Furthermore we will assume that $\widetilde{E}_{1,2}=v_{1,2}\, k+\kappa_{1,2}\, k^{2}$
(with no higher order corrections). We will assume that $\kappa_{-}=v_{+}=0$
and the density of states is set to $N_{0}={1}/{2\pi}$. We also scale
all units such that all quantities become dimensionless. We consider
an initial state ($t=0$) where the populations and coherences are
initialized at their zero-temperature equilibrium values $\widetilde{n}_{\kk}^{21}=0$,
$\widetilde{n}_{\kk}^{11}=n_{F}\left(E_{1}(\kk)\right)$ and $\widetilde{n}_{\kk}^{22}=n_{F}\left(E_{2}(\kk)\right)$.
The superconducting correlations are initialized at a very small but
non-zero value $\widetilde{s}_{\kk}^{21}=0.02\cdot\widetilde{s}_{\kk,Eq}^{21}$
where $\widetilde{s}_{\kk,Eq}^{21}$ is the steady state anomalous
correlator as computed in Sect.~\ref{sec:Mean-field-Hamiltonian}
(we also considered random initial conditions and obtained similar
results). We then time evolve the equations (\ref{eq:n11_rotated}),
(\ref{eq:n22_rotated}), (\ref{eq:n21_rotated}) and (\ref{eq:s21_rotated})
until we reach a steady state. We have used the self-consistency relation
in Eq.~(\ref{eq:self_consistency_corrected}). In Fig.~\ref{fig:Good_match_plots},
we present our numeric simulations for three representative coupling
constants (where $\Omega\gg\Gamma$ and $\gamma_{1}\gg\gamma_{2}$).
In Fig.~\ref{fig:Good_match_plots}(a), we plot the superconducting
gap as a function of time: it converges to the order parameter theoretically
predicted in Eq.~(\ref{eq:delta_large_omega}). To get good matching
we have calculated the correction to Eq.~(\ref{eq:delta_large_omega})
due to the finite cutoff in $k$-space $\left|k_{{\rm max}}\right|=0.2$
used in the numerical simulations. In Fig.~\ref{fig:Good_match_plots}(b)
we plot the theoretically predicted values of the anomalous correlator
${\widetilde{s}}_{\kk}^{21}$ as a function of $\kk$. We generate
${\widetilde{s}}_{\kk}^{21}$ in two different ways: one using the
theoretical predictions for the steady state given in Eqs.~(\ref{eq:Delta-steady-state})
and (\ref{eq:population}) and using the value of $\left|\Delta\right|$
from the simulations and also using the final values of the anomalous
correlators ${\widetilde{s}}_{\kk}^{21}$ as computed from the numerical
integrations. The agreement is excellent.

\begin{figure}
\begin{centering}
\hspace{1.5cm}\includegraphics[bb=100bp 0bp 360bp 252bp,scale=0.72]{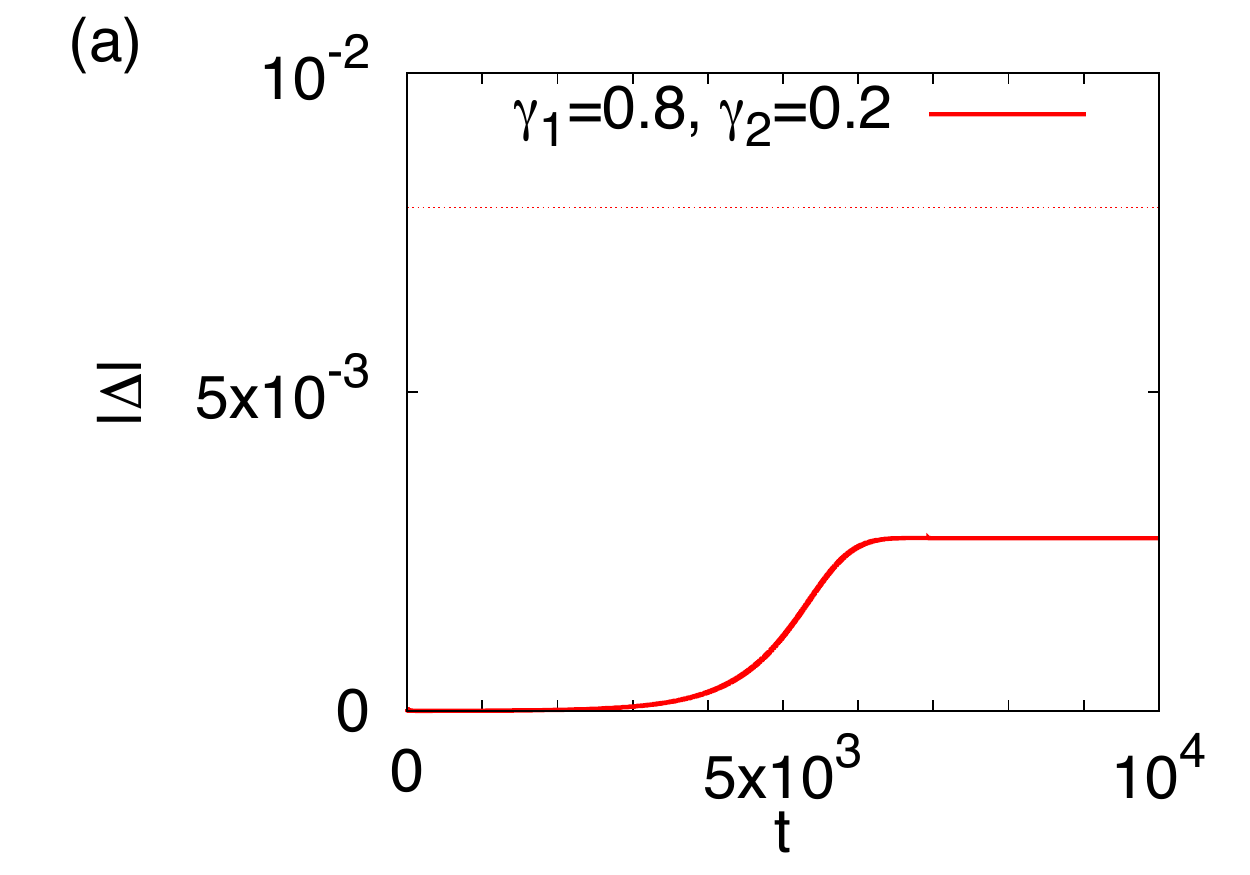}
\vspace{4.5cm}

\par\end{centering}

\begin{centering}
\hspace{-2.5cm}\includegraphics[bb=100bp 0bp 360bp 252bp,scale=0.4]{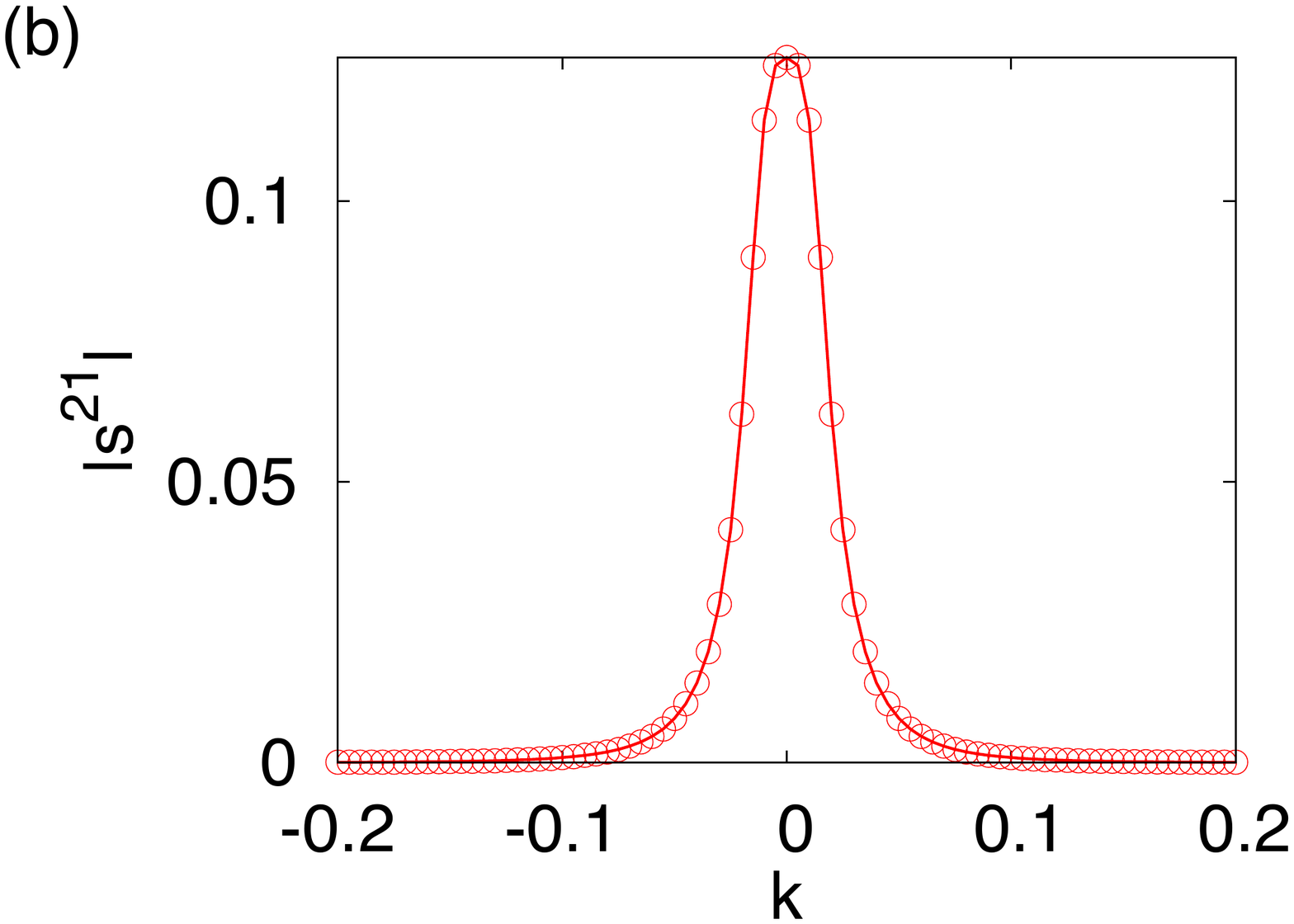}
\vspace{-2.5cm}

\par\end{centering}

\protect \protect\protect\protect\protect\protect\protect\protect\protect\caption{\label{fig:Bad_matching} (a) Time evolution of the superconducting
pairing $\Delta$ for a scenario where $\gamma_{1}\sim\gamma_{2}$.
The straight line corresponds to the steady-state value computed with
Eq.~(\ref{eq:delta_large_omega}). The discrepancy between analytics
and numerics is because the parameters for the numeric integration
are outside the limits of the approximations used in Sect.~\ref{sec:Self-Consistency-Equation}.
(b) Perfect matching between the steady-state anomalous correlator
${\widetilde{s}}_{\kk}^{21}$ as given by Eq.~(\ref{eq:Delta-steady-state})
(straight lines) and the anomalous correlator ${\widetilde{s}}_{\kk}^{21}$
obtained numerically after the time dynamics have converged (circles).
($\kappa_{+}=-50$, $v_{-}=10$, $\Omega=0.5$, $\Gamma=10^{-2}$,
$\gamma_{1}=0.8$ and $V=5$). We note that in a typical semiconductor
$\Gamma\approx10^{-2}eV$. Using this value for $\Gamma$ we obtain
that time is measured in units of $6.6\times10^{-16}$s and $\Delta$
is measured in units of 1eV. Also for typical semiconductors $v_{-}=10^{-2}c$,
so the momentum is measured in unites of $10^{3}eV/c\approx5.3\cdot10^{25}kg\cdot m/s$.
Furthermore using these values the coupling constant $V$ is measured
in unites of $0.14N_{0}^{-1}$ (where $N_{0}$ is the density of states
at the surface ${\cal S}_{\omega_{0}}$).}
\end{figure}

To show that our theory is able to predict superconducting pairing
even when our approximations for $\widetilde{n}_{\kk}^{11}+\widetilde{n}_{\kk}^{22}\!-\!1$
and hence ${\widetilde{s}}_{\kk}^{21}$ are not accurate, see Eq.~(\ref{eq:deviation_large_omega}),
we have chosen parameters outside the approximations of Sect.~\ref{sec:Self-Consistency-Equation}.
One way to make these approximations inaccurate is to consider a value
of $\gamma_{1},\gamma_{2}\succeq\frac{\Omega^{2}}{\Gamma^{2}+\left|v_{-}\right|^{2}{\Gamma}/{\kappa_{+}}}$
see the discussion below Eq.~(\ref{eq:deviation_large_omega}). In
Fig.~\ref{fig:Bad_matching}~(a), we have plotted the value of $\left|\Delta\right|$
as a function of time, we see that despite the failure of Eq.~(\ref{eq:deviation_large_omega})
we still obtain a relatively strong superconducting paring within
a factor of three of the analytical one. We also have compared the
In Fig.~\ref{fig:Bad_matching}~(b) we plot the theoretically predicted
values of the anomalous correlator ${\widetilde{s}}_{\kk}^{21}$ as
a function of $\kk$. We generate ${\widetilde{s}}_{\kk}^{21}$ in
two different ways: one using the theoretical predictions for the
steady state given in Eqs.~(\ref{eq:Delta-steady-state}) and (\ref{eq:population})
and the value of $\left|\Delta\right|$ from the simulations and also
using the final values of the anomalous correlators ${\widetilde{s}}_{\kk}^{21}$
as computed from the numerical integrations. The agreement is excellent.
We conclude that the system also reaches a non-trivial steady state
for parameter ranges outside the validity of the approximations used
in Sect.~\ref{sec:Self-Consistency-Equation}.

We have also numerically verified that it is possible to obtain superconductivity
for the case when $\Omega<\Gamma$. We have numerically integrated
the time evolution of the order parameter for two such values of $\Omega$
and $\Gamma$. We chose $\gamma_{1}\gamma_{2}\sim0$ in order to have
a non-zero order parameter (see the discussion in Sect.~\ref{sub:Weak-Rabi-frequency}).
We see that the order parameter develops but the time evolution is
highly oscillatory and the time scale for convergence is increased
by $\sim100$. This is because one of the decay rates, $\Gamma_{2}$,
is very small so it takes a long time for the oscillations to decay.

\begin{figure}
\begin{centering}
\vspace{4.5cm}
\hspace{-4.6cm}\includegraphics[bb=1cm 1cm 10cm 10cm,scale=0.4]{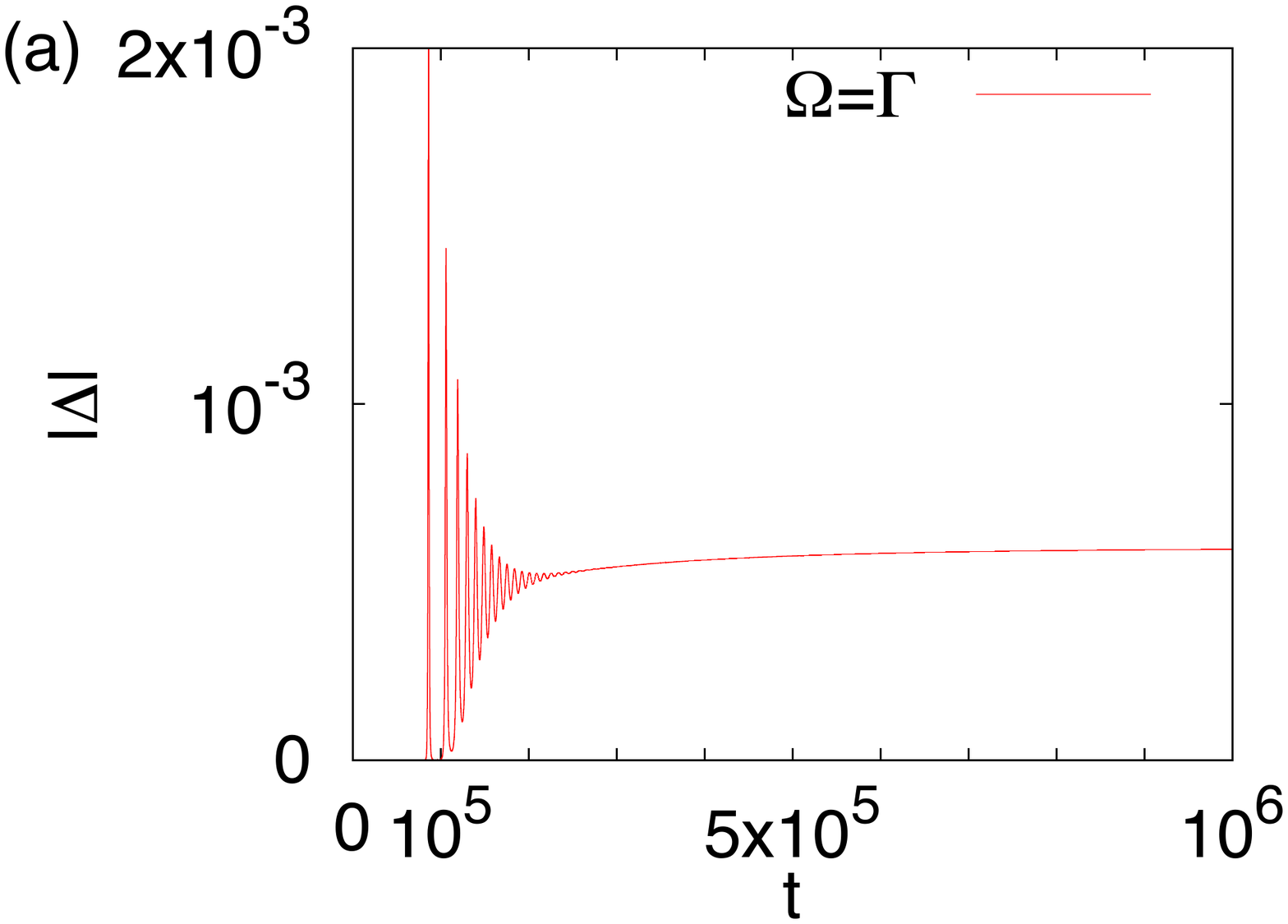}\vspace{-2.5cm}

\par\end{centering}

\begin{centering}
\hspace{2cm}\includegraphics[bb=100bp 0bp 360bp 252bp,scale=0.682]{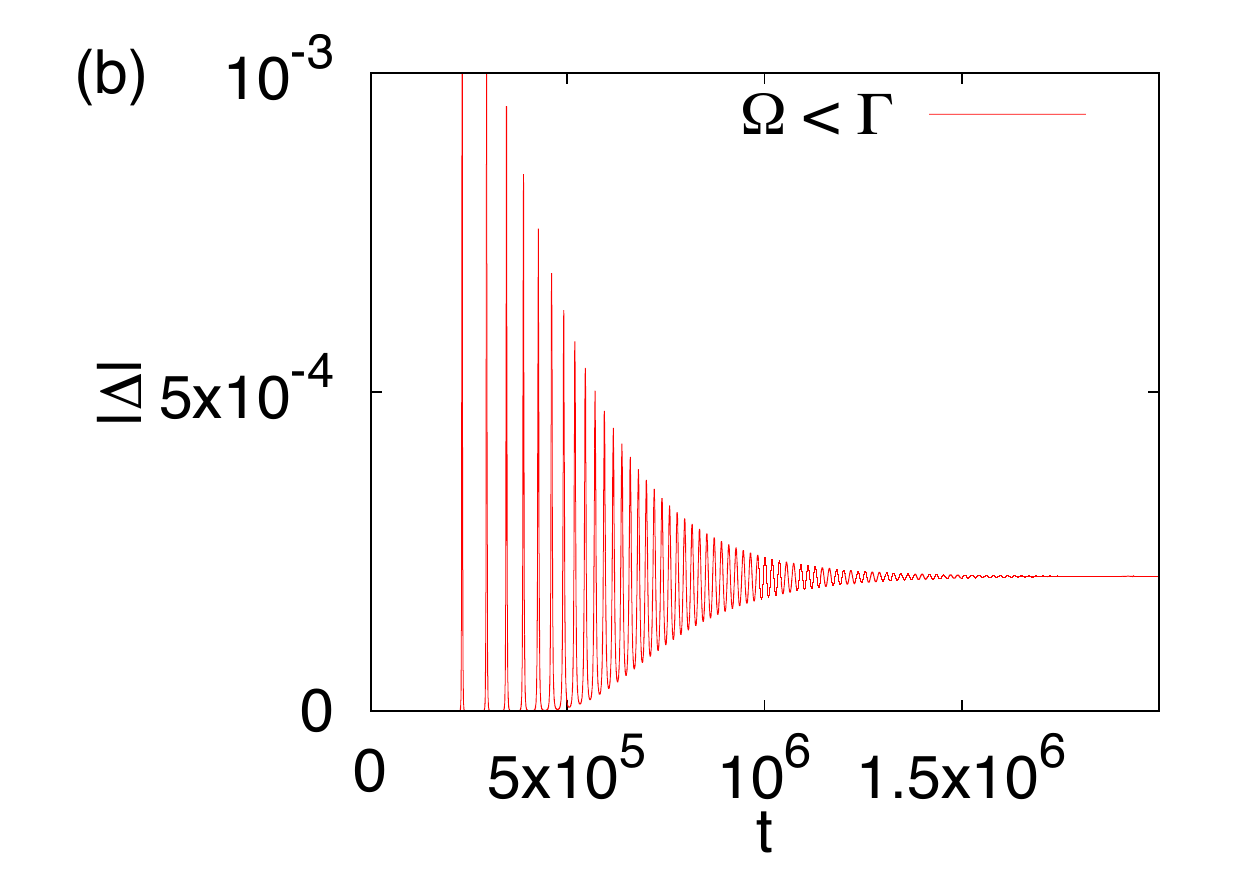} 
\par\end{centering}

\protect \protect\protect\protect\protect\protect\protect\protect\protect\caption{\label{fig:Small_omega}(a) Time evolution of the order parameter
$\Delta$ for $\Omega=\Gamma=10^{-2}$. (b) Time evolution of the
order parameter $\Delta$ for $\Omega=\Gamma/5=0.2\,10^{-2}$. ($\kappa_{+}=-50$,
$v_{-}=10$, $\Gamma=10^{-2}$, $\gamma_{2}=10^{-4}$, $V=20$). We
note that in a typical semiconductor $\Gamma\approx10^{-2}eV$. Using
this value for $\Gamma$ we obtain that time is measured in units
of $6.6\times10^{-16}$s and $\Delta$ is measured in units of 1eV.
Also for typical semiconductors $v_{-}=10^{-2}c$, using these values
the coupling constant $V$ is measured in unites of $0.14N_{0}^{-1}$
(where $N_{0}$ is the density of states at the surface ${\cal S}_{\omega_{0}}$).}
\end{figure}

\section{\label{sec:Optical-Pumping}Optical pumping of a two-band semiconductor}

Let us now turn to an alternate scenario, which may be more easily
realized in the lab. Let us consider a two-band semiconductor model
whose population of the bottom band is optically pumped into the upper
band \textit{via} a \emph{broad band} light source and whose interband
relaxation is slow, \textit{e.g.} negligible optical phonon coupling.
The lower band ($\alpha=1$) with dispersion $E_{1}(\kk)$ and the
upper band ($\alpha=2$) with dispersion $E_{2}(\kk)$ are separated
by a gap $E_{\rmg}$. 

\begin{figure}
\begin{centering}
\includegraphics[scale=0.3]{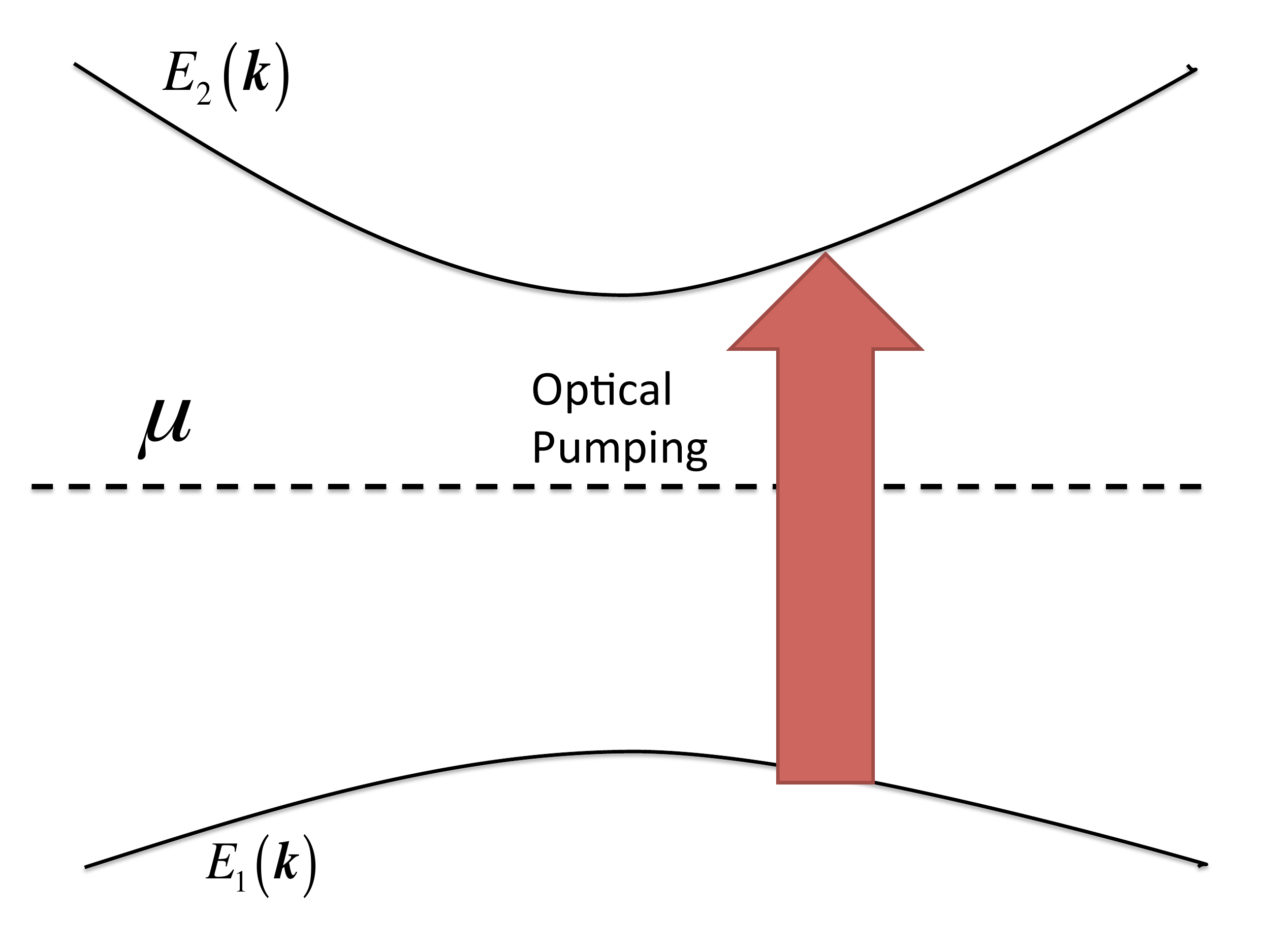} 
\par\end{centering}

\protect\protect\protect\protect\protect\protect\protect\protect\protect\caption{\label{fig:Optical_Pumping} Optical pumping. The upper band (1) of
a two-band semiconductor is populated with a single broad band optical
pump. The chemical potential $\mu$ is tuned halfway between the two
bands. }
\end{figure}

In order to reach a non-trivial steady state, the coupling to a thermal
reservoir is necessary to drain the energy which is continuously injected
in the system. However, unlike the previous case, the reservoir does
not need to play the role of an extra ``storage'' of particles (or
holes) and a single weakly-coupled reservoir is enough. We set the
chemical potential $\mu$ in the gap, see~Fig.~(\ref{fig:Optical_Pumping}),
such that there are momenta $\kk_{0}$ lying on a closed surface ${\cal S}$
of the Brillouin zone where the condition $E_{1}\left(\kk_{0}\right)+E_{2}\left(-\kk_{0}\right)=0$
is satisfied. Here, $\mu$ corresponds to the field produced by the
external voltages (say set by external gates).\textbf{ }We do not
assume that $E_{\alpha}\left(\kk_{0}\right)=\mbox{const}$. We shall
also assume the the optical pumping laser (or broadband source) is
not on resonance with these momenta $\kk_{0}$.

Neglecting superconductivity temporarily, the main effect of the optical
pumping is to modify the population of the lower and upper bands to
some non-trivial distribution. Since the pumping and the interband
relaxation is weak, the populations of the two bands relax to a separate
quasi-thermal equilibrium within each band. Therefore, the bands can
effectively be seen as having two different chemical potentials $\mu_{1}$
and $\mu_{2}$~\cite{key-32}. We note that $\mu_{1}$ and $\mu_{2}$
are not directly related to the energy levels of the Hamiltonian describing
the semiconductor. They can be seen as the Lagrange multipliers enforcing
the average number of particles in the two bands and depend on the
balance between the strength of the drive and the inter-band relaxation.
Once the system is quasi-equilibrated, we may write 
\begin{align}
n_{\kk}^{11} & =n_{{\rm F}}\left(E_{1}\left(\kk\right),\mu_{1}\right)\;,\nonumber \\
n_{-\kk}^{22} & =n_{{\rm F}}\left(E_{2}\left(-\kk\right),\mu_{2}\right)\;,\label{eq:Populations}\\
n_{\kk}^{12} & =0\;.\nonumber 
\end{align}
Here, $n_{{\rm F}}\left(\epsilon,\mu\right)\equiv[1+\exp\left(\left(\epsilon-\mu\right)/T\right)]^{-1}$
is the Fermi-Dirac distribution and $T$ is the temperature of the
underlying crystal. The equations of motion for the populations and
anomalous correlators which are consistent with the steady state given
in Eqs.~(\ref{eq:Populations}) read: 
\begin{align}
\frac{\rmd}{\rmd t}n_{\kk}^{11}= & \rmi\Delta s_{\kk}^{21}-\rmi\Delta^{\ast}s_{\kk}^{21\ast}-2\Gamma_{1}\left(\kk\right)\tilde{n}_{\kk}^{1}\nonumber \\
\frac{\rmd}{\rmd t}n_{-\kk}^{22}= & \rmi\Delta s_{\kk}^{21}-\rmi\Delta^{\ast}s_{\kk}^{21\ast}-2\Gamma_{2}\left(\kk\right)\tilde{n}_{-\kk}^{2}\label{eq:Population_equations_pumping}\\
\frac{\rmd}{\rmd t}s_{\kk}^{21}= & \rmi\left(E_{\kk}\left(\kk\right)-\rmi\Gamma_{21}\left(\kk\right)\right)s_{\kk}^{21}+\rmi\Delta^{\ast}\left(n_{\kk}^{11}+n_{-\kk}^{22}-1\right)\nonumber 
\end{align}
Here, $E_{\kk}\equiv E_{1}\left(\kk\right)+E_{2}\left(\kk\right)$
and $\tilde{n}_{\kk}^{\alpha}\equiv\left[n_{\kk}^{\alpha\alpha}-n_{F}\left(E_{\alpha}\left(\kk\right),\mu_{\alpha}\right)\right]$.
$\Gamma_{\alpha}$ are the relaxation rates for the two bands and
$\Gamma_{12}$ is the superconducting decay rate. In principle, these
can be obtained by linearizing the Boltzmann equation (collision integral)
close to equilibrium. Typically, $\Gamma_{12}\propto\Gamma_{1}+\Gamma_{2}$
\cite{key-32}. We also drop the $\kk$ dependence of $\Gamma$ since
we are only considering a small portion of the Brillouin zone near
the surface ${\cal S}$. The steady-state solution of these equations
reads 
\begin{equation}
{s}_{\kk}^{21}=-\frac{\Delta^{*}}{E_{\kk}+\rmi\Gamma_{12}}\;\left({n}_{\kk}^{11}+{n}_{-\kk}^{22}-1\right)\,\label{eq:Delta-steady-state-1}
\end{equation}
and 
\begin{align}
n_{\kk}^{11}+n_{-\kk}^{22}-1= & \frac{1}{\Xi'}\;4\gamma_{1}\gamma_{2}\,\left({\Gamma}/{\Gamma_{12}}\right)^{2}\\
 & \times\left[n_{{\rm F}}\left(E_{1}(\kk),\mu_{1}\right)+n_{{\rm F}}\left(E_{2}(-\kk),\mu_{2}\right)-1\right],\nonumber 
\end{align}
where we defined 
\begin{equation}
\Xi'\equiv4\gamma_{1}\gamma_{2}\,(\Gamma/\Gamma_{12})^{2}+\frac{4|\Delta|^{2}}{{E_{\kk}}^{2}+\Gamma_{12}^{2}}\;.
\end{equation}

\paragraph{Other pumping schemes}

\begin{figure}
\begin{centering}
\includegraphics[angle=90,scale=0.3]{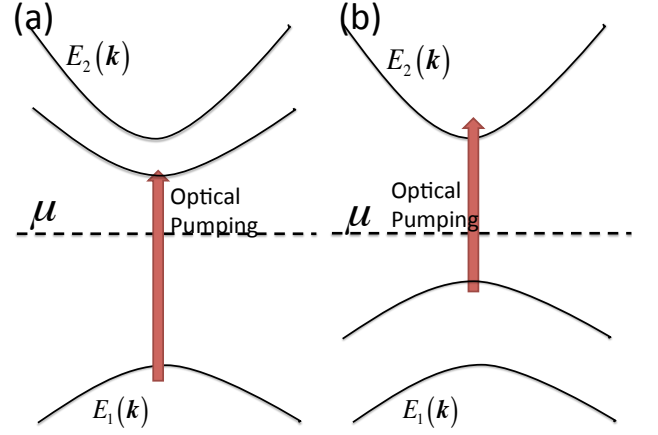} 
\par\end{centering}

\protect\protect\protect\protect\protect\protect\protect\protect\protect\caption{\label{fig:Optical_Pumping-1} Different available optical pumping
mechanisms in a three band semiconductor. (a) Deplete the population
of the bottom band into a third reservoir band. (b) Populate the top
band from a third band.}
\end{figure}

If other bands are present, other pumping schemes can be considered.
For instance, a third band can be used to either populate or depopulate
the two other bands, see Fig.~(\ref{fig:Optical_Pumping-1}). We
note that with these pumping schemes we can choose the sign of the
population deviation $n_{\kk}^{11}-n_{-\kk}^{22}-1$. Also, our method
is likely to work with carrier injection pumping~\cite{key-40}.
The conclusions presented in this Section apply just as well for these
generalized scenarios.

\subsection{\label{sub:Self-consistency-equation}Self-consistency equation}

We now solve self-consistently for the superconducting gap. The pairing
part of the mean-field Hamiltonian originates from a microscopic Hamiltonian
which involves a density-density type of interaction between the electrons
in the semiconductor. The mean-field decoupling for this microscopic
interaction of strength $V$ (in a system of volume ${\cal V}$) is
given by: 
\begin{eqnarray}
H_{e-e} & = & \frac{1}{{\cal V}}\sum_{\kk,\kk'}\; V\;{c_{\mathbf{k}}^{2}}^{\dagger}\,{c_{-\mathbf{k}}^{1}}^{\!\!\!\dagger}\;\;{c_{\mathbf{k}'}^{1}}\,{c_{-\mathbf{k}'}^{2}}\label{eq:superconductivity-1}\\
\qquad & \rightarrow & \sum_{\kk}\left(\Delta\;{c_{\mathbf{k}}^{2}}^{\dagger}\,{c_{-\mathbf{k}}^{1}}^{\!\!\!\dagger}+\Delta^{*}\;{c_{\mathbf{k}}^{1}}\,{c_{-\mathbf{k}}^{2}}\right)\;,\nonumber 
\end{eqnarray}
with 
\begin{equation}
\Delta^{*}=\frac{1}{{\cal V}}\sum_{\kk}V\langle{c_{\mathbf{k}}^{2}}^{\dagger}{c_{-\mathbf{k}}^{1}}^{\!\!\!\dagger}\rangle\xrightarrow[{\cal V}\to\infty]{}\int(\rmd{\kk})V\langle{c_{\mathbf{k}}^{2}}^{\dagger}{c_{-\mathbf{k}}^{1}}^{\!\!\!\dagger}\rangle\;,\label{eq:self-consistent-conditionb}
\end{equation}
where we wrote $(\rmd\kk)\equiv\rmd^{d}\kk/(2\pi)^{d}$ to shorten
notations.

We solve for the self-consistent condition Eq.~(\ref{eq:self-consistent-conditionb})
using the anomalous correlator in Eq.~(\ref{eq:Delta-steady-state-1}).
The correct self-consistent condition involves only the real part
of Eq.~(\ref{eq:Delta-steady-state-1}); this assertion will be justified
in Sect~\ref{sec:Keldysh-Calculation} where we properly obtain the
self-consistency relation from a saddle point condition (notice that
this is trivially true in the limit $\Gamma\to0$). The resulting
gap equation is 
\begin{eqnarray}
1 & = & -V\;\int(\rmd\kk)\;\frac{\gamma_{1}\gamma_{2}\,\left({\Gamma}/{\Gamma_{12}}\right)^{2}E_{\kk}}{\gamma_{1}\gamma_{2}\,\left({\Gamma}/{\Gamma_{12}}\right)^{2}\left(E_{\kk}^{2}+\Gamma_{12}^{2}\right)+|\Delta|^{2}}\nonumber \\
 &  & \times\left[n_{{\rm F}}\left(E_{1}(\kk),\mu_{1}\right)+n_{{\rm F}}\left(E_{2}(-\kk),\mu_{2}\right)-1\right]\;.\label{eq:gap-equation-1b}
\end{eqnarray}
Let us now study the solutions of the self-consistent equation (\ref{eq:gap-equation-1b})
by first focusing on the very favorable case in which the two bands
have opposite velocities. On the resonant surface ${\cal S}$, where
$E_{1}(\kk)+E_{2}\left(-\kk\right)=0$, the dispersion relations can
be Taylor-expanded as $\widetilde{E}_{1,2}=v_{1,2}\, q_{\perp}+\kappa_{1,2}\, q_{\perp}^{2}+\dots$,
where $q_{\perp}$ is the momentum perpendicular to the resonant surface
${\cal S}$. So $E=v_{+}\, q_{\perp}+\kappa_{+}\, q_{\perp}^{2}+\dots$,
where $v_{\pm}=v_{2}\pm v_{1}$ and $\kappa_{\pm}=\kappa_{2}\pm\kappa_{1}$.
When the velocities are opposite in the two bands, \textit{i.e.} $v_{+}=0$,
one can express $E(\epsilon)\approx(\kappa_{+}/v_{-}^{2})\;\epsilon^{2}$.
Upon using this $E(\epsilon)$ in Eq.~(\ref{eq:gap-equation-1b})
and extending the limits of integration to $\pm\infty$, we obtain:
\begin{align}
1= & -\frac{\pi}{\sqrt{2}}N_{0}V\,\,\frac{|v_{-}|\,{\rm sgn}\,\kappa_{+}}{\sqrt{|\kappa_{+}|}}\,\left(\gamma_{1}\gamma_{2}\right)^{1/4}\,\left(\Gamma/\Gamma_{12}\right)^{1/2}\nonumber \\
 & \times\frac{\left[n_{{\rm F}}\left(E_{1}(\kk),\mu_{1}\right)+n_{{\rm F}}\left(E_{2}(-\kk),\mu_{2}\right)-1\right]}{(|\Delta|^{2}+\gamma_{1}\gamma_{2}\Gamma^{2})^{1/4}}\;.\label{eq:self_consistency_optical_pumping}
\end{align}
Here $N_{0}$ is the density of states at ${\cal S}$. We note that
in the case where $\kappa_{+}$ is not uniform over the surface $ $${\cal S}$
we can replace $\sqrt{\left|\kappa_{+}\right|}$ in the equation above
by its average to obtain the correct results for this case. We will
not consider this extension further. Notice that this equation can
be satisfied for \emph{both attractive or repulsive interactions}
depending on the relative signs of $n_{{\rm F}}\left(E_{1}(\kk),\mu_{1}\right)+n_{{\rm F}}\left(E_{2}(-\kk),\mu_{1}\right)-1$
and of $\kappa_{+}$. Superconductivity is possible if the sign of
$V$ satisfies 
\begin{align}
{\rm sgn}\, V= & {\rm -sgn}\,\kappa_{+}\label{eq:optical_pumping_superconductivity_signs}\\
 & \times{\rm sgn}\left[n_{{\rm F}}\left((E_{1}(\kk),\mu_{1}\right)+n_{{\rm F}}\left(E_{2}(-\kk),\mu_{2}\right)-1\right]\;,\nonumber 
\end{align}
and if its magnitude satisfies the threshold condition 
\begin{eqnarray}
|V|\ge V_{\rmc}^{'''}\equiv\frac{\sqrt{2}}{\pi}\frac{1}{N_{0}}\,\frac{\sqrt{|\kappa_{+}|}}{\mathrm{N}|v_{-}|}\,\sqrt{\Gamma_{12}}\;.\label{eq:threshold-1}
\end{eqnarray}
Here $\mathrm{N}\equiv n_{{\rm F}}\left(E_{1}(\kk),\mu_{1}\right)+n_{{\rm F}}\left(E_{2}(-\kk),\mu_{2}\right)-1$.
The condition in Eq.~(\ref{eq:threshold-1}) is very similar to the
one obtained in Eq.~(\ref{eq:threshold}). This expresses the fact
that superconductivity is favored by small decay rates, \textit{e.g.}
weak coupling to longitudinal phonons and impurities.

If the conditions in Eqs.~(\ref{eq:optical_pumping_superconductivity_signs})
and (\ref{eq:threshold-1}) are met, the superconducting gap is given
by 
\begin{eqnarray}
|\Delta|=\sqrt{\gamma_{1}\gamma_{2}}\;\Gamma\;\sqrt{\left(\frac{V}{V_{\rmc}^{'''}}\right)^{4}-1}\;.
\end{eqnarray}
This corresponds to a robust gap that scales linearly with the decay
rate $\Gamma$, and, for large coupling constant, scales as the square
of the interaction strength $V$.

\paragraph{Robustness.}

Let us examine the domain of validity of the results we presented
in this Section. First, we remark that they are relatively stable
in the case $v_{+}$ is non-vanishing. Indeed, the results are essentially
unchanged as long as 
\begin{equation}
\left|v_{+}\right|\leq\sqrt{\left|\kappa_{+}\right|\Gamma}\,\frac{\left|V\right|}{V_{\rmc}^{'''}}\;.\label{eq:v+inequality}
\end{equation}
Therefore, the condition $v_{+}=0$ that we used above does not have
to be perfectly tuned.

Most importantly, the results of these Section are stable to changes
of temperature in the semiconductor. Indeed those would correspond
to changes in $n_{{\rm F}}(E_{1}(\kk))$ and $n_{{\rm F}}(E_{2}(\kk))$
which may be neglected for temperatures less then the semiconducting
gap $E_{\rmg}$. We note that in realistic setups, $\Gamma$ may be
temperature dependent.

\section{\label{sec:Keldysh-Calculation}Keldysh approach}

\begin{figure}
\begin{centering}
\includegraphics[angle=90,scale=0.3]{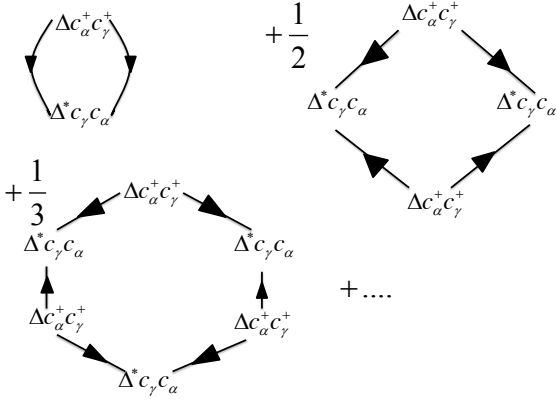} 
\par\end{centering}

\protect\protect\protect\protect\protect\protect\protect\protect\protect\caption{\label{fig:Feyman_diagrams} Feynman diagrams. Series of ring diagrams
that contribute to the action $\widetilde{S}[\Delta]$. Each line
corresponds to a propagator $G^{A/R/K}$. The terms $1/n$ are symmetry
factors for the diagrams. }
\end{figure}

In this Section, we revisit the self-consistent mean-field condition
for superconductivity that we used multiple times in the previous
Sections. Starting from a particle conserving theory, we justify the
approximation that we used to obtain Eqs.~(\ref{eq:gap-equation})
and (\ref{eq:gap-equation-1b}) which consisted in considering only
the part of the anomalous correlator $s_{\kk}^{\dagger21}$ in phase
with $\Delta^{*}$. For the sake of simplicity, we concentrate on
the case described in Sect.~\ref{sec:Mean-field-Hamiltonian}. We
derive a Keldysh mean-field theory for the laser-driven semiconductor
system and solve for the symmetry-breaking order-parameter corresponding
to the superconducting pairing. The full Keldysh action reads \begin{subequations}
\begin{equation}
S_{{\rm K}}=S_{e-e}+S_{{\rm other}}
\end{equation}
with 
\begin{equation}
S_{e-e}=\int_{\Upsilon}\rmd t\int\rmd^{d}\rr\; V\;\bar{\Phi}\left(\rr,t\right)\Phi\left(\rr,t\right)\;,\label{eq:Keldysh_Action}
\end{equation}
\end{subequations} where $\Phi(\rr,t)\equiv c^{1}(\rr,t)\, c^{2}(\rr,t)$,
$\bar{\Phi}(\rr,t)=c^{2\dagger}(\rr,t)c^{1\dagger}(\rr,t)$, $V$
is the coupling strength and $S_{{\rm other}}$ is the quadratic action
corresponding to all the other terms in the Hamiltonian~(\ref{eq:Main_Hamiltonian})
such as $c_{\kk}^{1}$, $c_{\kk}^{2}$, $a_{\kk,n}^{1}$ and $a_{\kk,n}^{2}$.
$\Upsilon$ is the Keldysh contour which goes forward from time minus
infinity to plus infinity and then backward. We now perform a Hubbard-Stratonovich
transformation in $S_{e-e}$ so as to obtain \begin{widetext} 
\begin{eqnarray}
\exp\left[{\rmi\int_{\Upsilon}\rmd t\int\rmd^{d}\rr\; V\;\bar{\Phi}\left(\rr,t\right)\Phi\left(\rr,t\right)}\right]=\int\mathcal{D}[\Delta]\;\;\rme^{\rmi\int_{\Upsilon}\rmd t\int\rmd^{d}\rr\;\left[-\frac{1}{V}\left|\Delta(\rr,t)\right|^{2}+\Delta\left(\rr,t\right)\,\bar{\Phi}\left(\rr,t\right)+\Delta^{*}\left(\rr,t\right)\,{\Phi}\left(\rr,t\right)\right]}\;.\label{eq:Hubbard_Stratonovich}
\end{eqnarray}
\end{widetext} Integrating out all the fields in $S_{{\rm K}}$ except
for $\Delta$, we obtain an effective action for $\Delta\left(\rr,t\right)$
and the zero-source generating functional reads 
\begin{eqnarray}
Z=\int\mathcal{D}[\Delta^{+},\Delta^{-}]\;\;\rme^{\rmi S_{{\rm eff}}[\Delta^{+},\Delta^{-}]}\;,\label{eq:Z_eff}
\end{eqnarray}
with the effective action expressed in terms of the fields $\Delta^{+}$
and $\Delta^{-}$ which correspond to the order-parameter in the forward
and backward branch of the Keldysh contour 
\begin{align}
S_{{\rm eff}}[\Delta^{+},\Delta^{-}]\equiv & \widetilde{S}[\Delta^{+},\Delta^{-}]\label{eq:S_eff}\\
 & -\frac{1}{V}\int\rmd t\rmd^{d}\rr\left(|\Delta^{+}(\rr,t)|^{2}-|\Delta^{-}(\rr,t)|^{2}\right).\nonumber 
\end{align}
$\widetilde{S}[\Delta^{+},\Delta^{-}]$ can be computed through a
series of Feynman diagrams as represented in Fig.~(\ref{fig:Feyman_diagrams}).
The propagators for these diagrams are those that make for the action
$S_{{\rm other}}$. Given that $S_{{\rm other}}$ is Gaussian, we
use Wick's theorem to calculate those Feynman diagrams.

We solve for the saddle point of the effective action by focusing
on the solutions that are homogeneous in time and space. We write
$\Delta^{\pm}=\Delta\pm\delta$, and note that the effective action
vanishes for $\delta=0$ for any $\Delta$. This is a general result
that stems from the fact that for classical field configurations,
the action on the backward branch is canceled exactly by that of the
forward branch. Thus, the variation of the effective action with respect
to $\Delta$ vanish for fixed $\delta=0$. The condition that determines
$\Delta$ at the saddle, is obtained by varying the action with respect
to $\delta$: expanding the action in powers of $\delta$, the saddle
point condition is that the terms linear in $\delta$ vanish. These
terms can be collected in perturbation theory.

Expanding $S_{{\rm eff}}[\Delta,\delta]$, we observe that all the
terms contain $\delta\Delta^{*}$, $\delta^{*}\Delta$, and powers
of $|\Delta|^{2}$. The action is invariant under simultaneous phase
rotations of $\delta$ and $\Delta$. So we can fix the phase of $\delta$
to be zero, \textit{i.e.} make $\delta$ real (this is, of course,
a gauge choice for the fermionic description of the problem). All
terms linear in $\delta$ are multiplying the combination $(\Delta+\Delta^{*})$
and powers of $|\Delta|^{2}$. Factoring out this combination $\delta(\Delta+\Delta^{*})$
in the expansion of $S_{{\rm eff}}[\Delta,\delta]$ leads to an equation
that depends only on $|\Delta|$. This equation determines the saddle
point value for $|\Delta|$. We choose $\Delta$ to be real as well,
and then simplify the saddle point search by considering both $\delta$
and $\Delta$ in phase and real. The net effect of this procedure
is to neglect the relative phase fluctuations of $\Delta^{+}$ and
$\Delta^{-}$ -- which are assumed to be small for a physical solution.
The saddle point equation in this case becomes 
\begin{equation}
0=\partial_{\delta}\;\left[-\frac{4\delta\Delta}{V}+\widetilde{{\cal L}}\left[\Delta,\delta\right]\right]_{\delta=0}\;.\label{eq:Gap_Equation}
\end{equation}

We compute $\partial_{\delta}\widetilde{{\cal L}}\left[\Delta,\delta\right]\big|_{\delta=0}$
by summing over the Feynman diagrams in Fig.~(\ref{fig:Feyman_diagrams})
and obtain \begin{widetext} 
\begin{align}
 & \partial_{\delta}\widetilde{{\cal L}}\left[\Delta,\delta\right]\big|_{\delta=0}=\\
 & \sum_{n=1}^{\infty}\left|\frac{\Delta}{2}\right|^{2n-1}\!\!\!\int(\rmd\kk)\int\frac{\rmd\omega}{2\pi}\;{\rm Tr}\;\Bigg\{\begin{pmatrix}0 & \openone\\
\openone & 0
\end{pmatrix}\left[\begin{pmatrix}G^{R}(\kk,\omega) & G^{K}(\kk,\omega)\\
0 & G^{A}(\kk,\omega)
\end{pmatrix}\begin{pmatrix}\rmi\sigma^{y}G^{A}(\kk,-\omega)\rmi\sigma^{y} & \rmi\sigma^{y}G^{K}(\kk,-\omega)\rmi\sigma^{y}\\
0 & \rmi\sigma^{y}G^{R}(\kk,-\omega)\rmi\sigma^{y}
\end{pmatrix}\right]^{n}\Bigg\}\nonumber \\
 & +\sum_{n=1}^{\infty}\left|\frac{\Delta}{2}\right|^{2n-1}\!\!\!\int(\rmd\kk)\int\frac{\rmd\omega}{2\pi}\;{\rm Tr}\;\Bigg\{\begin{pmatrix}0 & \openone\\
\openone & 0
\end{pmatrix}\left[\begin{pmatrix}\rmi\sigma^{y}G^{A}(\kk,-\omega)\rmi\sigma^{y} & \rmi\sigma^{y}G^{K}(\kk,-\omega)\rmi\sigma^{y}\\
0 & \rmi\sigma^{y}G^{R}(\kk,-\omega)\rmi\sigma^{y}
\end{pmatrix}\begin{pmatrix}G^{R}(\kk,\omega) & G^{K}(\kk,\omega)\\
0 & G^{A}(\kk,\omega)
\end{pmatrix}\right]^{n}\Bigg\}\;.\nonumber 
\end{align}
Here $G^{A/R/K}$ stand for the advanced, retarded and Keldysh components
of the electronic Green's functions for the bands $1$ and $2$, with
respect to the action $S_{{\rm other}}$, and $\sigma^{y}$ is the
usual Pauli matrix which acts on the space spanned by the two bands
$\alpha=1,\,2$. (Notice that $G^{A/R/K}$ are $2\times2$ matrices
because of the two bands.) The Pauli matrix $\sigma^{y}$ and the
negative frequencies $-\omega$ in some of the Green's functions come
about because some of the propagators shown in Fig.~(\ref{fig:Feyman_diagrams})
originate from the same vertex (or, equivalently, there are particle
and hole propagators). We now observe that this series can be resumed
and the trace can be greatly simplified: 
\begin{align}
\partial_{\delta}\widetilde{{\cal L}}\left[\Delta,\delta\right]\big|_{\delta=0} & =\frac{\left|\Delta\right|}{4\pi}\int\left(\rmd\kk\right)\int\rmd\omega\mbox{ Tr}\left[\left(1-\frac{\left|\Delta\right|^{2}}{4}G^{R}\left(\kk,\omega\right)\rmi\sigma_{y}G^{A}\left(\kk,-\omega\right)\rmi\sigma_{y}\right)^{-1}\right.\nonumber \\
 & \qquad\times\left(G^{R}\left(\kk,\omega\right)\rmi\sigma_{y}G^{K}\left(\kk,-\omega\right)\rmi\sigma_{y}+G^{K}\left(\kk,\omega\right)\rmi\sigma_{y}G^{R}\left(\kk,-\omega\right)\rmi\sigma_{y}\right)\nonumber \\
 & \qquad\left.\times\left(1+\left(1-\frac{\left|\Delta\right|^{2}}{4}G^{A}\left(\kk,\omega\right)\rmi\sigma_{y}G^{R}\left(\kk,-\omega\right)\rmi\sigma_{y}\right)^{-1}\frac{\left|\Delta\right|^{2}}{4}G^{A}\left(\kk,\omega\right)\rmi\sigma_{y}G^{R}\left(\kk,-\omega\right)\rmi\sigma_{y}\right)\right]\nonumber \\
 & \quad+\frac{\left|\Delta\right|}{4\pi}\int\left(\rmd\kk\right)\int\rmd\omega\mbox{ Tr}\left[\left(1-\frac{\left|\Delta\right|^{2}}{4}\rmi\sigma_{y}G^{A}\left(\kk,-\omega\right)\rmi\sigma_{y}G^{R}\left(\kk,\omega\right)\right)^{-1}\right.\nonumber \\
 & \qquad\times\left(\rmi\sigma_{y}G^{K}\left(\kk,-\omega\right)\rmi\sigma_{y}G^{A}\left(\kk,\omega\right)+\rmi\sigma_{y}G^{A}\left(\kk,-\omega\right)\rmi\sigma_{y}G^{K}\left(\kk,\omega\right)\right)\nonumber \\
 & \qquad\left.\times\left(1+\left(1-\frac{\left|\Delta\right|^{2}}{4}\rmi\sigma_{y}G^{R}\left(\kk,-\omega\right)\rmi\sigma_{y}G^{A}\left(\kk,\omega\right)\right)^{-1}\frac{\left|\Delta\right|^{2}}{4}\rmi\sigma_{y}G^{R}\left(\kk,-\omega\right)\rmi\sigma_{y}G^{A}\left(\kk,\omega\right)\right)\right]\;.\label{eq:steady_state}
\end{align}
In case the superconducting field $\left|\Delta\right|$ is small,
this expression Eq.~(\ref{eq:steady_state}) may be further simplified:
\begin{align}
\approx & \frac{\left|\Delta\right|}{2}\int(\rmd\kk)\,\int\frac{\rmd\omega}{2\pi}\;{\rm Tr}\;\left[\rmi\sigma_{y}G^{K}\left(\kk,-\omega\right)\rmi\sigma_{y}G^{A}\left(\kk,\omega\right)+\rmi\sigma_{y}G^{A}\left(\kk,-\omega\right)\rmi\sigma_{y}G^{K}\left(\kk,\omega\right)\right.\nonumber \\
 & \qquad\qquad\qquad\qquad\left.+G^{R}\left(\kk,\omega\right)\rmi\sigma_{y}G^{K}\left(\kk,-\omega\right)\rmi\sigma_{y}+G^{K}\left(\kk,\omega\right)\rmi\sigma_{y}G^{R}\left(\kk,-\omega\right)\rmi\sigma_{y}\right]\nonumber \\
 & +\;\frac{\;\left|\Delta\right|^{3}}{8}\int(\rmd\kk)\,\int\frac{\rmd\omega}{2\pi}\;{\rm Tr}\;\left[\left(\rmi\sigma_{y}G^{K}\left(\kk,-\omega\right)\rmi\sigma_{y}G^{A}\left(\kk,\omega\right)+\rmi\sigma_{y}G^{A}\left(\kk,-\omega\right)\rmi\sigma_{y}G^{K}\left(\kk,\omega\right)\right)\right.\nonumber \\
 & \qquad\qquad\qquad\qquad\left.\times\left(\rmi\sigma_{y}G^{A}\left(\kk,-\omega\right)\rmi\sigma_{y}G^{R}\left(\kk,\omega\right)+\rmi\sigma_{y}G^{R}\left(\kk,-\omega\right)\rmi\sigma_{y}G^{A}\left(\kk,\omega\right)\right)\right]\nonumber \\
 & +\frac{\;\left|\Delta\right|^{3}}{8}\int(\rmd\kk)\,\int\frac{\rmd\omega}{2\pi}\;{\rm Tr}\;\left[\left(G^{R}\left(\kk,\omega\right)\rmi\sigma_{y}G^{K}\left(\kk,-\omega\right)\rmi\sigma_{y}+G^{K}\left(\kk,\omega\right)\rmi\sigma_{y}G^{R}\left(\kk,-\omega\right)\rmi\sigma_{y}\right)\right.\nonumber \\
 & \qquad\qquad\qquad\qquad\left.\times\left(G^{R}\left(\kk,\omega\right)\rmi\sigma_{y}G^{A}\left(\kk,-\omega\right)\rmi\sigma_{y}+G^{A}\left(\kk,\omega\right)\rmi\sigma_{y}G^{R}\left(\kk,-\omega\right)\rmi\sigma_{y}\right)\right]\;.\label{eq:Equilibrium}
\end{align}
Using the quantum regression theorem~\cite{key-30}, one can compute
the various Green's functions $G^{R}(\kk,\omega)=(\omega-H(\kk)+\rmi\hat{\Gamma})^{-1}$,
$G^{A}(\kk,\omega)=(\omega-H(\kk)-\rmi\hat{\Gamma})^{-1}$ and $G^{K}(\kk,\omega)=G^{R}(\kk,\omega)\left(1-2f(\kk)\right)-(1-2f(\kk))G^{A}(\kk,\omega)$.
Here $H(\kk)\equiv\left(\begin{array}{ll}
\widetilde{E}_{1}(\kk) & \Omega/2\\
\Omega/2 & \widetilde{E}_{2}(\kk)
\end{array}\right)$, $\hat{\Gamma}\equiv\left(\begin{array}{ll}
\Gamma_{1} & 0\\
0 & \Gamma_{2}
\end{array}\right)$ and $f\left(\kk\right)\equiv\left(\begin{array}{ll}
n_{\kk}^{11} & n_{\kk}^{21}\\
n_{\kk}^{12} & n_{\kk}^{22}
\end{array}\right)$. We may now perform the various traces and integrals over $\omega$
in Eq.~(\ref{eq:Equilibrium}) above. With this, we solve for the
stationary conditions on the field $\Delta$, coming from Eq.~(\ref{eq:Gap_Equation}),
and obtain 
\begin{equation}
0=\frac{\left|\Delta\right|}{V}+\left|\Delta\right|\int(\rmd\kk)\;{\rm Re}\left\{ \frac{1}{{E}_{\kk}+\rmi\Gamma}\right\} \;(1-n_{\kk}^{11}-n_{\kk}^{22})-4\left|\Delta\right|^{3}\int(\rmd\kk)\;{\rm Re}\left\{ \frac{1}{{E}_{\kk}+\rmi\Gamma}\right\} \;\frac{1-n_{\kk}^{11}-n_{\kk}^{22}}{{E}_{\kk}^{2}+\Gamma^{2}}\;.\label{eq:Consistency_relation}
\end{equation}
\end{widetext} The part involving ${\rm Re}\left\{ \frac{1}{{E}_{\kk}+\rmi\Gamma}\right\} $
is exact and comes about because the Keldysh action must be real.
In the third term of Eq.~(\ref{eq:Consistency_relation}), we have
also made the assumption that $\Omega\gg\Gamma$. Using the non-equilibrium
population deviation $1-n_{\kk}^{11}-n_{\kk}^{22}$ given in Eq.~(\ref{eq:deviation_large_omega})
for small $\left|\Delta\right|$ we see that this agrees to leading
order for small $\left|\Delta\right|$ with Eq.~(\ref{eq:gap-equation});
a computation of the exact trace in Eq.~(\ref{eq:steady_state})
would presumably reproduce Eq.~(\ref{eq:gap-equation}) to all orders.

\section{\textup{\label{sec:Conclusions}Conclusions} }

We have demonstrated that superconductivity can be achieved in a laser-driven
two-band semiconductor interacting with reservoir -- either in the
form of a tunneling contact to a metal, or in the form of other modes
in the band, or in the form of a third band (see appendix). The superconductivity
is robust to changes in temperature, and under optimal conditions,
the size of the superconducting gap scales with the decay rate $\Gamma$.
We found that depending on the sign of the band curvatures, it is
possible to obtain superconducting pairing $s_{\kk}^{21}$ with both
repulsive and attractive interactions. We can estimate how stringent
is the condition given in Eq.~(\ref{eq:threshold}) for the threshold
for producing superconductivity with two bands and reservoirs. To
do so we compare our results to regular BCS theory. At zero temperature,
the BCS gap equation may be written as $V\rho\left(k_{{\rm F}}\right)\ln\left(\frac{\omega_{{\rm D}}}{\Delta}\right)=1$.
Here $\rho\left(k_{{\rm F}}\right)$ is the density of states at Fermi
energy and $\omega_{{\rm D}}$ is the Debye frequency. Using the experimentally
relevant parameters $\omega_{{\rm D}}\sim100K$ and $\Delta\sim1K$
we obtain $V\rho\left(k_{{\rm F}}\right)\sim0.2$. We note that $V$
is the effective electron-electron interaction which includes the
effects of phonons and screening. For the superconductivity proposed
in this manuscript we have obtained the threshold equation $\frac{\left|V\right|\, N_{0}\left|v_{-}\right|}{\left|\kappa\right|^{1/2}\Gamma^{1/2}}>1$.
Using $\kappa\sim\left(10^{6}eV\right)^{-1}c^{2}$, $\Gamma\sim10^{-3}eV$,
$\left|v{}_{-}\right|\sim10^{-2}c$ this condition simplifies to $0.2\times10^{2.5}>1$
which is easily satisfied. We note that for the case of repulsive
interactions $\left|V\right|\, N_{0}$ can be larger. These numbers
are relevant for room temperature superconductivity. We note that
the same threshold condition shows up in the case of an optically
pumped two-band semiconductor considered in Sect.~\ref{sec:Optical-Pumping}
and in the case where the laser Rabi frequency is small but the two
decay rates are very different, see Eqs.~(\ref{eq:supercondcuting_treshhold})
and (\ref{eq:threshold-1}). Eq.~(\ref{eq:v+inequality}) establishes
that all our results are unaffected by mismatches in the Fermi velocities
of the upper and lower band as long as these mismatches are only roughly
ten percent of the Fermi velocity. The present results are also insensitive
to imprecision in tuning the right $\mu$ on the order of 0.01~$e$V.
We note that imperfections in finding the right $\mu$ do not effect
the results presented in Sect.~\ref{sec:Optical-Pumping} as the
condition $E_{1}\left(\kk\right)+E_{2}\left(-\kk\right)=0$ is automatically
selected. Even though $T_{\rmc}$ (critical temperature for superconductivity)
does not scale with the gap for our setup, as in the case of a regular
superconductor, we note that under optimal conditions it is possible
to achieve a gap that is several hundred Kelvins.

We unveiled a new route to induce superconductivity, not simply by
lowering the temperature of the sample but by shining light. In a
semiconductor, such photo-induced superconductivity is possible at
temperatures smaller than the band gap, which itself is a very high
temperature. Hence, the mechanism may enable dissipationless current
transport for frequencies smaller than that set by the superconducting
gap at room temperature. In many ways the ultimate limit on our setup
is the temperature dependence of the rate $\Gamma$. $T_{c}$ is set
by the relationship $\frac{\left|V\right|\, N_{0}\left|v_{-}\right|}{\left|\kappa\right|^{1/2}\Gamma^{1/2}\left(T_{c}\right)}=1$.
Additionally, one can imagine applications where the superconductivity
is induced for short periods of time by laser pulses and is allowed
to decay when the laser is turned off. This opens the door for superconducting
switches. We intend to perform a DMFT analysis of the phenomena to
study the effects of strong correlations and strong laser driving.
We shall also study the optical response of the proposed superconductor
as well as investigate the possibility of a Josephson effect.

\acknowledgments

This work has been supported by the Rutgers CMT fellowship (G.G.),
the NSF grants DMR-0906943 and DMR-115181 (C.A.), and the DOE Grant
DEF-06ER46316 (C.C.). We would like to thank L. Levitov, G. Kotliar
and D. Khmelnitskii for pointing out useful references.

\appendix


\section{\label{sec:System-Requirements}Three-band semi-conductor}

\begin{figure}
\begin{centering}
\includegraphics[angle=90,scale=0.3]{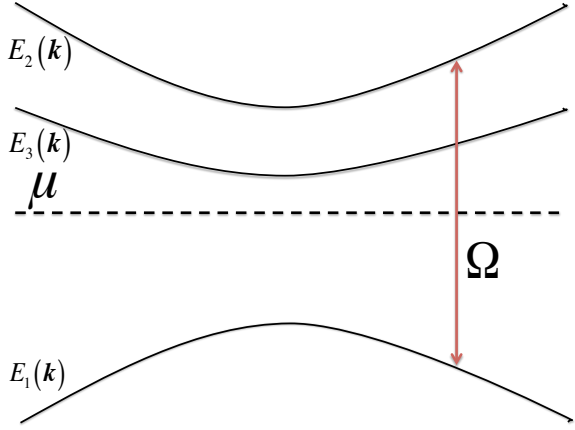} 
\par\end{centering}

\protect\protect\protect\protect\protect\caption{\label{fig:Energy_Levels} Energy levels and laser. The upper band
of a three-band semiconductor is populated with a single laser drive
pumping from the lower band. The chemical potential $\mu$ is set
between the lower and middle (reservoir) bands. }
\end{figure}

Let us consider superconductivity in an optically pumped three-band
semiconductor. The lower (1), middle (3) and upper (2) bands have
dispersion relations $E_{\alpha}(\kk)$ with $\alpha=1,\,2,\,3$.
The third band can be seen as a replacement for the reservoirs that
were required in the case considered in Sect.~\ref{sec:Mean-field-Hamiltonian}.
We assume that the dispersion is symmetric so that $E_{\alpha}\left(\kk\right)=E_{\alpha}\left(-\kk\right)$,
for $\alpha=1,\,2$ -- this will allow for s-wave superconductivity
without any energy mismatch. The upper and lower bands are resonantly
driven by a single laser with frequency $\omega_{0}$, \textit{i.e.}
$E_{2}\left(\kk_{0}\right)=E_{1}\left(\kk_{0}\right)+\omega_{0}$
for some wave vectors $\kk_{0}$. In our scheme, the middle band $3$
(reservoir) is not coupled to any laser but will simply ensure that
there is less then one electron per $\kk$ value in the upper and
lower bands combined, $n_{\kk}^{11}+n_{\kk}^{22}<1$. This inequality
satisfies the condition that the population of the two bands involved
in the pairing deviates from unity, which was the requisite for superconductivity
in Sect.~II. We set the chemical potential $\mu$ in between the
lower and middle band, \textit{i.e.} $E_{2}>E_{3}>\mu>E_{1}$ for
all wave vectors $\kk$, see Fig.~(\ref{fig:Energy_Levels}). More
precisely, we set $\mu=[E_{2}(\kk_{0}+E_{1}(\kk_{0})]/2$; this ensures
that all quasiparticles have zero energy. In the rotating frame, this
will correspond to a zero energy condition for the electrons at $\kk_{0}$.

To favor superconductivity, we assume that the level sets of $E_{1}\left(\kk_{0}\right)$
and $E_{2}\left(\kk_{0}\right)$ have a good overlap and that the
electron velocities of the lower and upper bands are opposite at the
wave vector $\kk_{0}$. Under such conditions, we find that depending
on the curvature of the lower and upper bands at $\kk_{0}$ it is
possible to induce superconductivity with \emph{either repulsive or
attractive interactions}, in particular to obtain a non-vanishing
anomalous correlator $\left\langle c_{\kk}^{2}c_{-\kk}^{1}\right\rangle \equiv s_{\kk}^{21}$.
The analysis presented in this Appendix is highly similar to the one
done in the body of the manuscript and will be presented briefly.

\subsection{\label{sec:Mean-field-Hamiltonian-1}Mean-field Hamiltonian}

The mean-field Hamiltonian relevant to our three-band system can be
written as: 
\begin{align}
H_{{\rm MF}}=H_{{\rm Band}}+H_{{\rm Laser}}+H_{{\rm Super}}\;,\label{eq:Hamiltonian_pieces}
\end{align}
with 
\begin{align}
H_{{\rm Band}}=\sum_{\kk,\alpha}E_{\alpha}\left(\kk\right)c_{\kk}^{\alpha\dagger}c_{\kk}^{\alpha}\;,\label{eq:H_band}\\
H_{{\rm Laser}}=\sum_{\kk}\Omega\left(t\right)c_{\kk}^{2\dagger}c_{\kk}^{1}+\mbox{ h.c.}\;,\label{eq:H_laser}\\
H_{{\rm Super}}=\sum_{\kk}\Delta c_{\kk}^{2\dagger}c_{-\kk}^{1\dagger}+\mbox{ h.c.}\;.\label{eq:H_super}
\end{align}
$\Omega\left(t\right)\equiv\Omega\cos\left(\omega_{0}t\right)$ is
the laser drive and $\Delta$ is the mean-field superconducting gap.
The relevant equation of motions read 
\begin{align}
\rmi\frac{\rmd}{\rmd t}c_{\kk}^{2}= & \Delta c_{-\kk}^{1\dagger}+E_{2}c_{\kk}^{2}+\Omega\left(t\right)c_{\kk}^{1}\;,\nonumber \\
\rmi\frac{\rmd}{\rmd t}c_{\kk}^{3}= & E_{3}c_{\kk}^{3}\;,\nonumber \\
\rmi\frac{\rmd}{\rmd t}c_{\kk}^{1}= & -\Delta c_{-\kk}^{2\dagger}+E_{1}c_{\kk}^{1}+\Omega^{*}\left(t\right)c_{\kk}^{2}\;.\label{eq:Equation_motion}
\end{align}
We eliminate all explicit time dependence by a means of rotating wave
approximation. This consists in rotating all the operators of the
theory with the unitary 
\begin{align}
U\equiv U_{c}\otimes U_{a}\;,
\end{align}
where 
\begin{align}
U_{c}\equiv & \exp\left[\frac{\rmi}{2}\omega_{0}t\sum_{\kk}\left(c_{\kk}^{1\dagger}c_{\kk}^{1}-c_{\kk}^{2\dagger}c_{\kk}^{2}\right)\right]\;,\\
U_{a}\equiv & \exp\left[\frac{\rmi}{2}\omega_{0}t\sum_{\kk,n}\left(a_{\kk,n}^{1\dagger}a_{\kk,n}^{1}-a_{\kk,n}^{2\dagger}a_{\kk,n}^{2}\right)\right]\;.
\end{align}
In particular, $c_{\kk}^{1}\mapsto\widetilde{c}_{\kk}^{1}=c_{\kk}^{1}\,\rme^{-\rmi\omega_{0}t/2}$,
$c_{\kk}^{2}\mapsto\widetilde{c}_{\kk}^{2}=c_{\kk}^{2}\,\rme^{\rmi\omega_{0}t/2}$,
and $H\mapsto\widetilde{H}=U\left[H-\rmi\partial_{t}\right]U^{\dagger}$
so that the energies are shifted to $\widetilde{E}_{1}(\kk)=E_{1}(\kk)+\omega_{0}/2$
and $\widetilde{E}_{2}(\kk)=E_{2}(\kk)-\omega_{0}/2$. We drop all
terms rotating at $2\omega_{0}$ since they are not resonant with
any transition. By adding dissipative mechanisms and considering fermion
bilinears we may study the non-equilibrium steady state properties
of this model. Note that in the rotating frame, $\widetilde{n}_{\kk}^{11}=n_{\kk}^{11}$,
$\widetilde{n}_{\kk}^{22}=n_{\kk}^{22}$, and $\widetilde{s}_{\kk}^{12}=s_{\kk}^{12}$
are invariant, but $\widetilde{n}_{\kk}^{12}=n_{\kk}^{12}\,\rme^{-\rmi\omega_{0}t}$
and $\widetilde{n}_{\kk}^{21}=n_{\kk}^{21}\,\rme^{\rmi\omega_{0}t}$.
The steady-state equation of the order-parameter $\widetilde{s}_{\kk}^{21}=\left\langle \widetilde{c}_{\kk}^{2}\widetilde{c}_{-\kk}^{1}\right\rangle $
reads \begin{widetext} 
\begin{equation}
0=\rmi\frac{\rmd}{\rmd t}\widetilde{s}_{\kk}^{\dagger21}=-\Delta^{*}(1-\widetilde{n}_{\kk}^{11}-\widetilde{n}_{\kk}^{22})-(\widetilde{E}_{\kk}-\rmi\Gamma_{12})\widetilde{s}_{\kk}^{\dagger21}\;,\label{eq:SteadyState}
\end{equation}
where we introduced the notation $\widetilde{E}_{\kk}\equiv\widetilde{E}_{1}(\kk)+\widetilde{E}_{2}(\kk)$
and $\Gamma_{12}$ is a phenomenological decay rate associated with
the damping of the order parameter. This simplifies to 
\begin{equation}
\widetilde{s}_{\kk}^{21}=\frac{\Delta^{*}}{\widetilde{E}_{\kk}+\rmi\Gamma_{12}}\left(\widetilde{n}_{\kk}^{11}+\widetilde{n}_{\kk}^{22}-1\right)\;.\label{eq:Final_value_supercondcutivity}
\end{equation}
This equation is identical to Eq.~(\ref{eq:Delta-steady-state})
and seems to be an ubiquitous condition for superconductivity. This
expresses that to ensure superconductivity, we once again need to
have a nonzero $\widetilde{n}_{\kk}^{11}+\widetilde{n}_{\kk}^{22}-1\neq0$.
This is the rationale behind the presence of the third band -- which
does not interact with the other two bands but merely acts as ``storage''
for electrons. To find the steady-state value of $\widetilde{n}_{\kk}^{11}$
and $\widetilde{n}_{\kk}^{22}$, we write the steady-state equations
for the rest of the fermion bilinears. From now on, we shall work
in the weak pairing field limit $\Delta\ll\Omega$. The steady-state
equations for the populations and coherences read 
\begin{align}
0=\frac{\rmd}{\rmd t}\widetilde{n}_{\kk}^{22}= & \rmi\frac{\Omega}{2}\left(\widetilde{n}_{\kk}^{12}-\widetilde{n}_{\kk}^{21}\right)-\left(\Gamma_{1}+\Gamma_{2}\right)\widetilde{n}_{\kk}^{22}\;,\nonumber \\
0=\frac{\rmd}{\rmd t}\widetilde{n}_{\kk}^{33}= & \Gamma_{1}n_{\kk}^{22}-\Gamma_{3}n_{\kk}^{33}\;,\nonumber \\
0=\frac{\rmd}{\rmd t}\widetilde{n}_{\kk}^{11}= & -\rmi\frac{\Omega}{2}\left(\widetilde{n}_{\kk}^{12}-n_{\kk}^{21}\right)+\Gamma_{2}\widetilde{n}_{\kk}^{22}+\Gamma_{3}\widetilde{n}_{\kk}^{33}\;,\nonumber \\
0=\frac{\rmd}{\rmd t}\widetilde{n}_{\kk}^{21}= & (\rmi\varepsilon_{\kk}-\tau^{-1})\widetilde{n}_{\kk}^{21}+\rmi\frac{\Omega}{2}\left(\widetilde{n}_{\kk}^{11}-\widetilde{n}_{\kk}^{22}\right)\;.\label{eq:SteadyStateMany}
\end{align}
\end{widetext} We have introduced three spontaneous decay rates $\Gamma_{1},\,\Gamma_{2},\,\Gamma_{3}$
and a dephasing time $\tau$. For many semiconductors $\tau^{-1}\gg\Gamma_{1,2,3}$
because it is hard to exchange populations between the bands, by including
say Coulomb interactions, but rather easy to have energy fluctuations
which lead to dephasing. In the case the semiconductor has a strong
coupling to optical phonons, this inequality may be violated as all
the decay rates may become comparable. Notice that the previous equations
ensure the conservation of particle number, \textit{i.e.} $\widetilde{n}_{\kk}^{11}+\widetilde{n}_{\kk}^{22}+\widetilde{n}_{\kk}^{33}=1$.
Using Eq.~(\ref{eq:Final_value_supercondcutivity}), we obtain 
\begin{equation}
s_{\kk}^{\dagger21}=-\frac{\Delta^{*}}{\widetilde{E}_{\kk}+\rmi\Gamma_{12}}\frac{\frac{\left|\Omega\right|^{2}\Gamma_{1}}{2\tau\left(\varepsilon_{\kk}^{2}+\tau^{-2}\right)}}{\Gamma_{3}\left(\Gamma_{1}+\Gamma_{2}\right)+\frac{\left|\Omega\right|^{2}\left(\Gamma_{1}+2\Gamma_{3}\right)}{2\tau\left(\varepsilon_{\kk}^{2}+\tau^{-2}\right)}}\;,\label{eq:Finalsuperconducting}
\end{equation}
with $\varepsilon_{\kk}\equiv\widetilde{E}_{2}(\kk)-\widetilde{E}_{1}(\kk)$.

Note that $s_{\kk}^{\dagger21}$ vanishes when $\Gamma_{1}=0$ but
$\Gamma_{3}\neq0$. In this case there is no population in the the
middle band, \textit{i.e.} $\widetilde{n}_{\kk}^{33}=0$. However,
one would not expect that $s_{\kk}^{\dagger21}=0$ if we simultaneously
tune $\Gamma_{1},\Gamma_{3}\downarrow0$ as some population will be
trapped in band $3$ (reservoir) if both decay rates go down to zero
with the same rate, which can be seen from the analysis of Eq.~(\ref{eq:Finalsuperconducting}).

\subsection{\label{sec:Self-Consistency-Equation-1}Self-Consistency Equation}

We now solve self-consistently for the superconducting gap. The pairing
part of the mean-field Hamiltonian in Eq.~(\ref{eq:Hamiltonian_pieces})
originates from a microscopic Hamiltonian which involves a density-density
type of interaction between the electrons in the semiconductor. The
corresponding mean-field decoupling is given in Eq.~(\ref{eq:superconductivity-1}).
To obtain most favorable conditions for superconductivity, we shall
once again assume that the electron velocities of the lower and upper
bands are opposite at the wave vector $\kk_{0}$. At the resonant
surface ${\cal S}_{\omega_{0}}$, the dispersion relation can be Taylor-expanded
as $\widetilde{E}_{1,2}=v_{1,2}\, q_{\perp}+\kappa_{1,2}\, q_{\perp}^{2}+\dots$,
where $q_{\perp}$ is the momentum perpendicular to the resonant surface
${\cal S}_{\omega_{0}}$ and $v_{1}+v_{2}=0$. So $\varepsilon=v_{-}\, q_{\perp}+\kappa_{-}\, q_{\perp}^{2}+\dots$
and $E=\kappa_{+}\, q_{\perp}^{2}+\dots$, where $v_{-}\equiv v_{2}-v_{1}$
and $\kappa_{\pm}=\kappa_{2}\pm\kappa_{1}$. Substituting these energies
into the gap equation, we obtain the condition \begin{widetext} 
\begin{equation}
\begin{array}{l}
\Delta^{*}\leq-VN_{0}\left|v_{-}\right|\,\int\rmd q_{\perp}\frac{\Delta^{*}\kappa_{+}q_{\perp}^{2}}{\kappa_{+}^{2}q_{\perp}^{4}+\Gamma_{12}^{2}}\;\frac{\frac{\left|\Omega\right|^{2}\Gamma_{1}}{2\tau\left(\left(v_{-}q_{\perp}\right)^{2}+\tau^{-2}\right)}}{\Gamma_{3}\left(\Gamma_{1}+\Gamma_{2}\right)+\frac{\left|\Omega\right|^{2}\left(\Gamma_{1}+2\Gamma_{3}\right)}{2\tau\left(\left(v_{-}q_{\perp}\right)^{2}+\tau^{-2}\right)}}\;.\end{array}\label{eq:self-consinstency}
\end{equation}
$N_{0}$ is the density of states at $\kk_{0}$. We note that 
\begin{equation}
{\rm sgn}\, V={\rm sgn}\,\kappa_{+}
\end{equation}
is needed to satisfy the condition, which means that by tuning band
curvatures it is possible to have superconductivity with \emph{both
attractive and repulsive interactions}. Furthermore, we note that
in the case in which the Rabi frequency is large, the integral in
Eq.~(\ref{eq:self-consinstency}) greatly simplifies and the threshold
condition for superconductivity becomes 
\begin{eqnarray}
|V|\ge V_{{\rm c}}\equiv\frac{\sqrt{2}}{\pi}\frac{1}{N_{0}}\frac{\sqrt{|\kappa_{+}|}}{|v_{-}|}\sqrt{\Gamma_{12}}\,\left(1+2\Gamma_{3}/\Gamma_{1}\right)\;.\label{eq:Self_Consistency_simplified}
\end{eqnarray}
\end{widetext} In the small damping limit (\textit{i.e.} small $\Gamma_{12}$),
the inequality is easily satisfied. This condition is highly similar
to the condition obtained for superconducting threshold in Eq.~(\ref{eq:threshold}).
We note that for the case $\gamma_{2}-\gamma_{1}\sim1$, $\Gamma_{12}\sim\Gamma$,
and $\Gamma_{1}\gg\Gamma_{3}$ the two equations become equivalent.


\begin{thebibliography}{10}
\bibitem{key-17} M. Tinkham, \textit{Introduction to superconductivity}
(Dover Publications, 1996).

\bibitem[2]{key-33} P. W. Anderson, \textit{The theory of superconductivity
in the high $T_{\rmc}$ cuprates}, (Princeton university press 1997).

\bibitem[3]{key-34} D. C. Johnston, Advances in Physics \textbf{59},
803 (2010).

\bibitem[4]{key-35} Q. Si and E. Abrahams, Phys. Rev. Lett. \textbf{101},
076401 (2008).

\bibitem[5]{key-36} J. E. Mooij, T. P. Orlando, L. Levitov, L. Tian,
C. H. van der Wal and S. Lloyd, Science \textbf{285}, 1036 (1999).

\bibitem[6]{key-37} J. Q. You, J. S. Tsai and F. Nori, Phys. Rev.
Lett. \textbf{89}, 197902 (2002).

\bibitem[7]{key-38}M. N. Wilson, \textit{Superconducting magnets}
(Clarendon press 1983).

\bibitem[8]{key-39} A. Mourachkine, \textit{Room temperature superconductivity}
(Cambridge International Science publishing 2004).

\bibitem[9]{key-18} D. H. Dunlap, and V. M. Kenkre, Phys. Rev. B
\textbf{34}, 3625 (1986).

\bibitem[10]{key-19} M. Holthaus, Phys. Rev. Lett. \textbf{69}, 351
(1992).

\bibitem[11]{key-20} H. Lignier, C. Sias, D. Ciampini, Y. Singh,
A. Zenesini, O. Morsch, and E. Arimondo, Phys. Rev. Lett. \textbf{99},
220403 (2007).

\bibitem[12]{key-21} C. E. Creffield and T. S. Monteiro, Phys. Rev.
Lett. \textbf{96}, 210403 (2006).

\bibitem[13]{key-22} A. Eckardt, C. Weiss, and M. Holthaus, Phys.
Rev. Lett. \textbf{95}, 260404 (2005).

\bibitem[14]{key-23} A. Zenesini, H. Lignier, D. Ciampini, O. Morsch,
and E. Arimondo, Phys. Rev. Lett. \textbf{102}, 100403 (2009).

\bibitem[15]{key-24} J. Struck, C. Ölschläger, R. Le Targat, P. Soltan-Panahi,
A. Eckardt, M. Lewenstein, P. Windpassinger, and K. Sengstock, Science
\textbf{333}, 996 (2011).

\bibitem[16]{key-25} N. Tsuji, T. Oka, P.Werner, and H. Aoki, Phys.
Rev. Lett. \textbf{106}, 236401 (2011).

\bibitem[17]{key-26} N. Tsuji, T. Oka, H. Aoki, and P. Werner, Phys.
Rev. B \textbf{85}, 155124 (2012).

\bibitem[18]{key-27} N. Tsuji, T. Oka, and H. Aoki, Phys. Rev. B
\textbf{78}, 235124 (2008).

\bibitem[19]{key-1} V. F. Elesin, Sov. Phys. JETP 32, 328 (1971).

\bibitem[20]{key-2} V. M. Galitskii, V. F. Elesin and Yu. V. Kopaev,
ZhETF Pis. Red. 18, 50 (1973).

\bibitem[21]{key-3} D. A. Kirshnits and Yu. V. Kopaev, ZhETF Pis.
Red. 17, 379 (1973).

\bibitem[22]{key-4} V. F. Elesin, Yu. V. Kopaev and R. Kh. Timerov,
Zh. Eksp. Teor. Fiz. 2343 (1973).

\bibitem[23]{key-5} V. M. Galitskii, S. P. Goreslavskii and V. F.
Elesin Zh. Eksp. Theor. Fiz. 57, 207 (1969).

\bibitem[24]{key-41} It is not strictly necessary that each wave
vector $k$ corresponds to its own reservoir. All that is needed is
that $\left\langle t^{\ast}\left(k\right)t\left(k'\right)\right\rangle \sim\delta\left(k-k'\right)$
which would happen for a disordered metal reservoir.

\bibitem[25]{key-1-1} R. R. Puri, \textit{Mathematical Methods of
Quantum Optics,} (Springer Verlag 2001).

\bibitem[26]{key-32} H. Haug and S. W. Koch, \textit{Quantum theory
of the optical and electronic properties of semiconductors}, (World
Scientific Publishing co. 1990).

\bibitem[27]{key-40} W. W. Chow, S. W. Koch and M. Sargent, \textit{Semiconductor-laser
physics} (Springer-Verlag 1994).

\bibitem[28]{key-30}H. G. Carmichael, \textit{Statistical Methods
in Quantum Optics 1: Master Equations and Fokker-Plank Equations},
(Springer Verlag 1999).\end{thebibliography}
\end{document}